\documentclass{scrartcl}

\usepackage{hhline}
\usepackage{multirow}
\usepackage{graphicx}
\usepackage{amsmath}
\usepackage{color}
\usepackage{longtable, ctable}
\usepackage{ctable}
\usepackage{array}
\usepackage[obeyspaces]{url}
\usepackage{stackengine}
\usepackage{units}
\usepackage{comment}

\newcommand\xrowht[2][0]{\addstackgap[.5\dimexpr#2\relax]{\vphantom{#1}}}
\newcolumntype{C}[1]{>{\centering\arraybackslash}p{#1}} 
\newcolumntype{R}[1]{>{\raggedleft\arraybackslash}p{#1}}                                                                  
\newcolumntype{L}[1]{>{\raggedright\arraybackslash}p{#1}}                                                                 

\newcommand{\um}{\unit{\mu m}}

\author{Melissa C. Revelle}
\title{Phoenix/Peregrine Trap}

\begin{document}
	\begin{center}\Huge{Phoenix and Peregrine Ion Traps}\end{center}
	
	\hrule
	\vspace{1cm}

	\begin{center}
		\includegraphics[width=\textwidth]{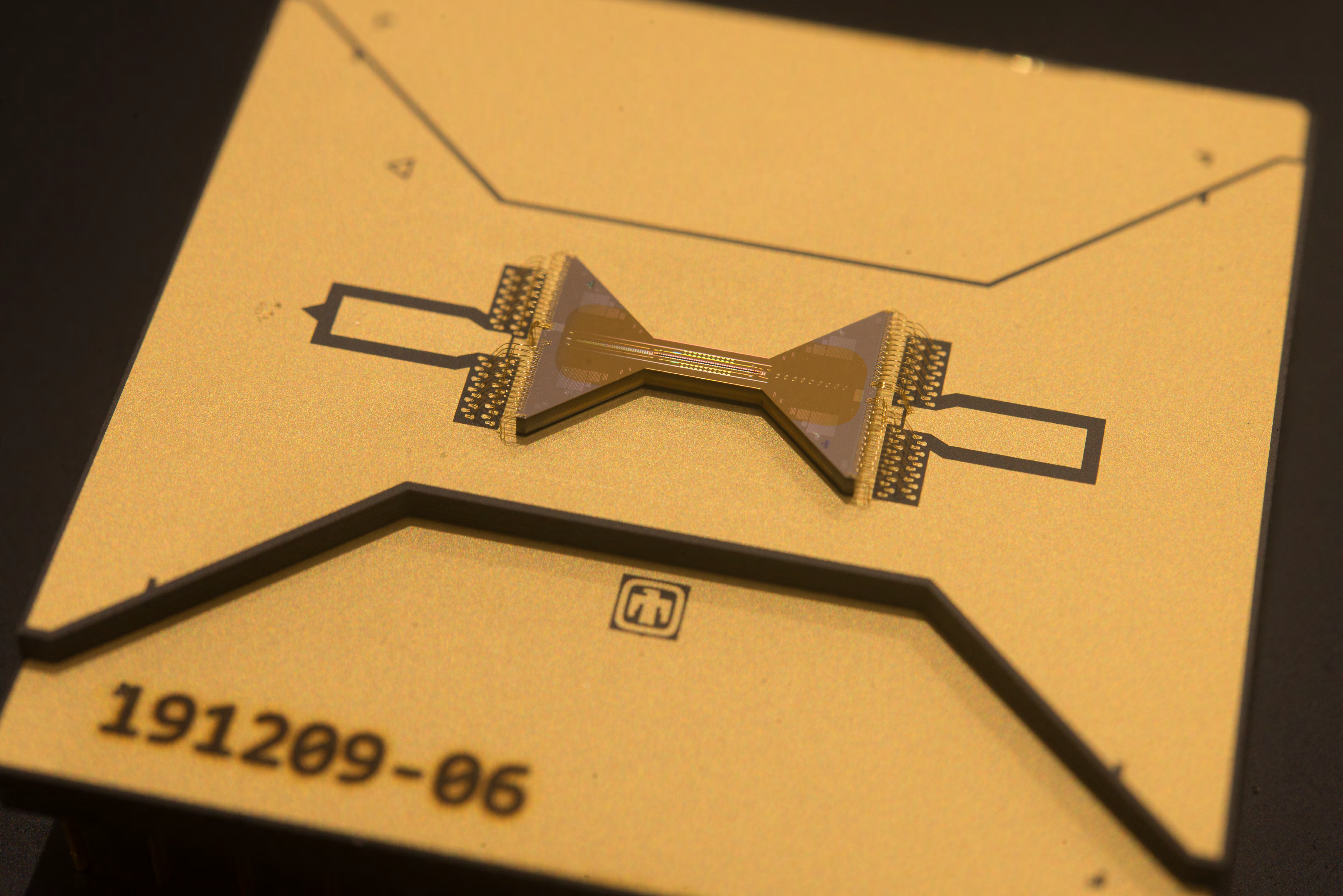}
		
		\vspace{1cm}
		{\Large{Version 1, \today}}
		
		\vspace{1cm}
		
		\large
		\em{Contact:}\\
		Melissa Revelle\\
		Sandia National Laboratories\\
		Building 858EL/3104; MS 1082\\
		1515 Eubank Blvd SE\\
		Albuquerque, NM  87123\\
		
		mrevell@sandia.gov\\
		505-844-9606
	\end{center}
	
	\newpage
	
	\parbox[c]{0.36\textwidth}{\includegraphics[width=0.36\textwidth]{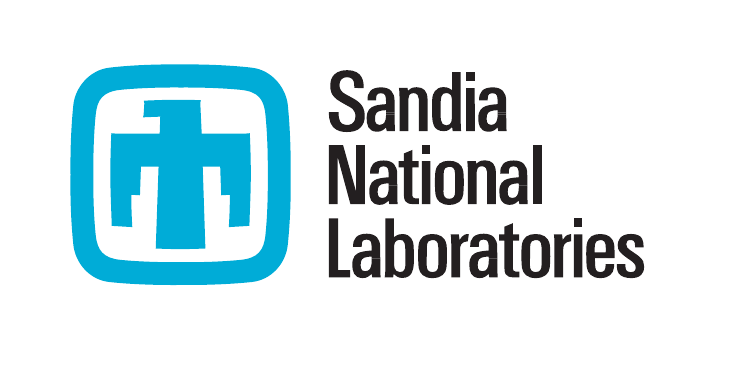}}
	\parbox[c]{0.6\textwidth}{\flushleft{
			Sandia National Laboratories is a multimission laboratory managed and operated by National Technology \& Engineering Solutions of Sandia, LLC, a wholly owned subsidiary of Honeywell International Inc., for the U.S. Department of Energy’s National Nuclear Security Adminis- tration under contract DE-NA0003525.
		}}
	\vspace{0.5cm}
	
	\parbox[c]{0.36\textwidth}{\includegraphics[width=0.3\textwidth]{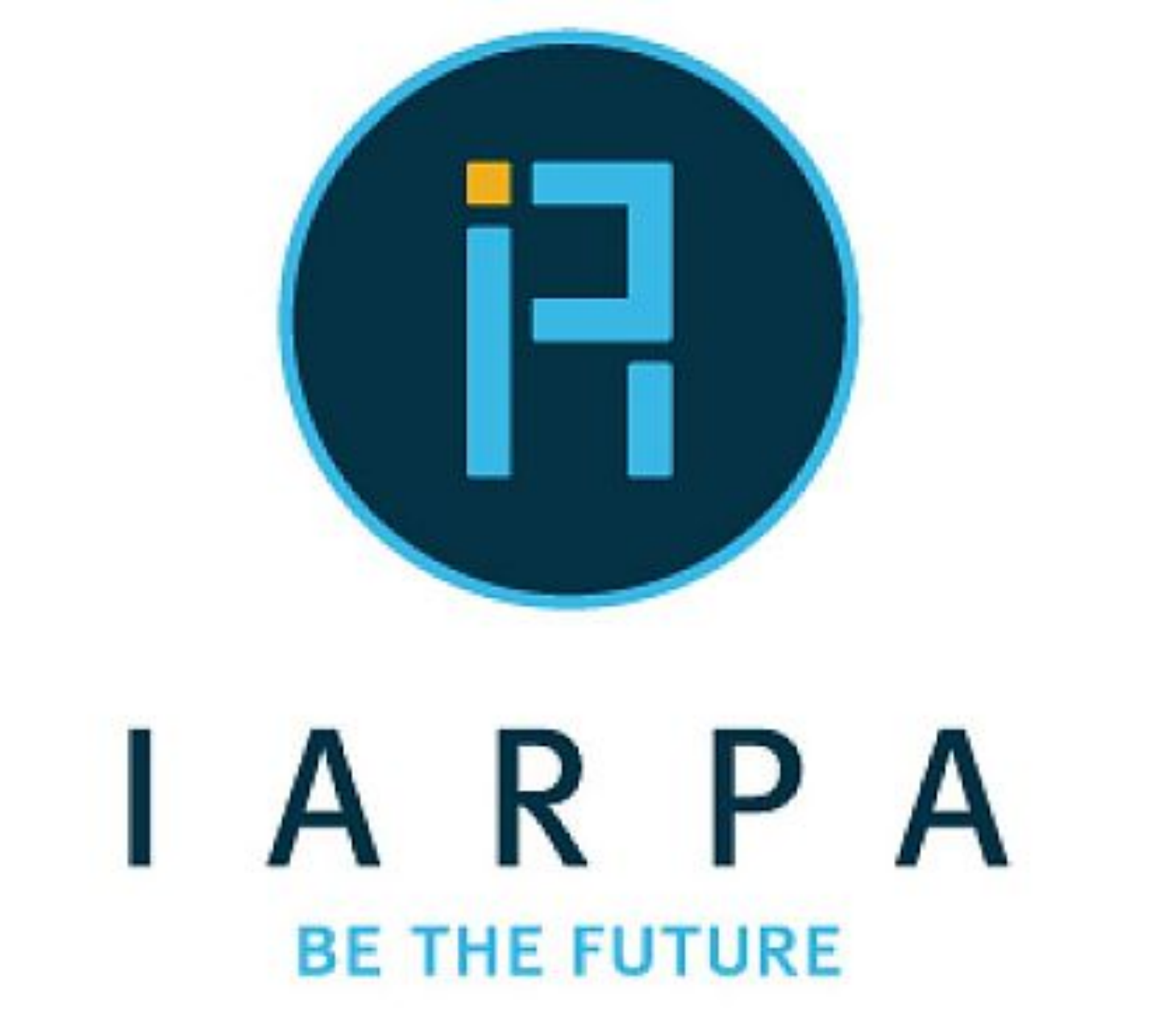}}
	\hfill
	\parbox[c]{0.6\textwidth}{\flushleft{This work was supported by the Intelligence Advanced Research Projects Activity (IARPA) under the Logical Qubits (LogiQ) program.}}
	
	\vspace{0.5cm}

	\parbox[c]{0.36\textwidth}{\includegraphics[width=0.35\textwidth]{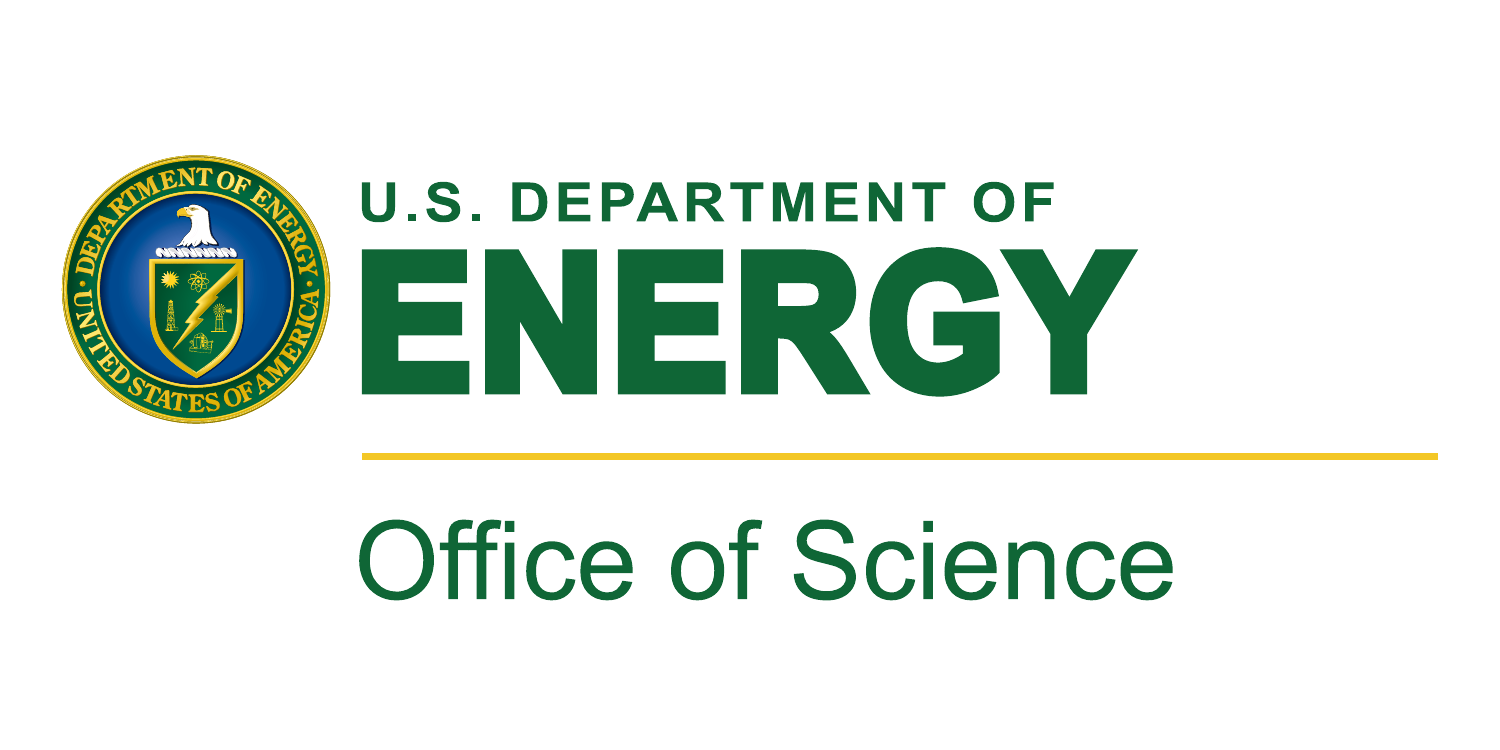}}
	\hfill
	\parbox[c]{0.6\textwidth}{\flushleft{And by the United States Department of Energy (DOE) Office of Science as part of the Advanced Scientific Computing Research (ASCR) Quantum Testbed program.}}

	\vfill

\newpage
\tableofcontents
\newpage
\begin{center}
	\textbf{\large Phoenix and Peregrine Microfabricated Surface Electrode Ion Traps\\[6pt]}

\end{center}

\begin{abstract}
	The Phoenix and Peregrine ion traps are micro-fabricated surface-electrode ion traps based on silicon technology.
	Both are linear traps using a symmetric 6-rail design with segmented inner and outer control electrodes. 
	The traps are fabricated on Sandia's High Optical Access (HOA) platform to provide good optical access skimming the trap surface.
	They are packaged in custom ceramic pin or land grid array packages using a $\unit[2.54]{mm}$ pitch. 
	The Peregrine trap is a surface trap with all electrodes in one plane. 
	The Phoenix trap has the same layout, but with a central through-substrate slot and its inner control electrodes are at a lower metal level.
	Both traps provide means to measure the substrate temperature and to heat the device by means of integrated aluminum and tungsten wires.
\end{abstract}

\section{Introduction}
The Phoenix and Peregrine traps are a set of related linear surface traps that have been developed at Sandia National Laboratories for IARPA's logical qubits (LogiQ) program and the Quantum Scientific Computing Open User Testbed (QSCOUT) project funded by the Department of Energy Office for Science, respectively.  

Microfabrication offers unique advantages for the fabrication of ion traps for applications in quantum information processing. 
Silicon CMOS (complementary metal-oxide-semiconductor) fabrication processes make it possible to fabricate traps that adhere precisely to the design. 
This makes it possible to not only produce identical traps, but also leads to very good predictability of the trapping potentials from electrostatic models.
Furthermore, a large number of control electrodes can be realized to precisely control trapping potentials.

Trap properties that are important for the use of ion traps in quantum information processing include the following:
\begin{itemize}
	\item High optical access for beams skimming the surface of the trap and if desired also for beams passing through a slot in the trap substrate.
	\item High radio-frequency (rf) voltage efficiency and sufficiently large radial trap frequencies to enable high-fidelity quantum gates.
	\item Full control of all degrees of freedom of the trapping potential.
	\item Integrated capacitors to ensure that control electrodes provide good rf ground.
	\item Good axial voltage efficiency to enable precise control of axial trapping fields as well as to realize separation and merging of ion chains.
	\item A loading region separated from the trap region used for quantum operations.
	\item Constant pseudo-potential magnitude throughout the entire trap to enable ion shuttling though the device while maintaining constant trapping conditions.
	\item Minimization of rf dissipation on the device to facilitate consistent trap operation and to reduce the heat load for operation at cryogenic temperatures.
\end{itemize}

In the design for the Phoenix and Peregrine traps we are building upon the experience of producing several generations of micro-fabricated surface electrode ion traps and are integrating solutions for requirements necessary to realize advanced quantum information processors.  
	
\section{Trap design}
The Phoenix and Peregrine traps are linear surface electrode ion traps built on Sandia's bowtie-shaped HOA platform~\cite{HOAManual}. 
The Phoenix trap includes a through-substrate slot that makes it possible to pass laser beams perpendicular to the trap surface and provides a numerical aperture of $0.25$. 
The Peregrine trap is a pure surface trap.\\
Both traps have a quantum region with independent, segmented inner and outer control electrodes and a loading hole for loading ions away from the quantum region. 

\subsection{Phoenix Trap}

\begin{figure}[h]
	\centering
	\includegraphics[width=1.0\textwidth]{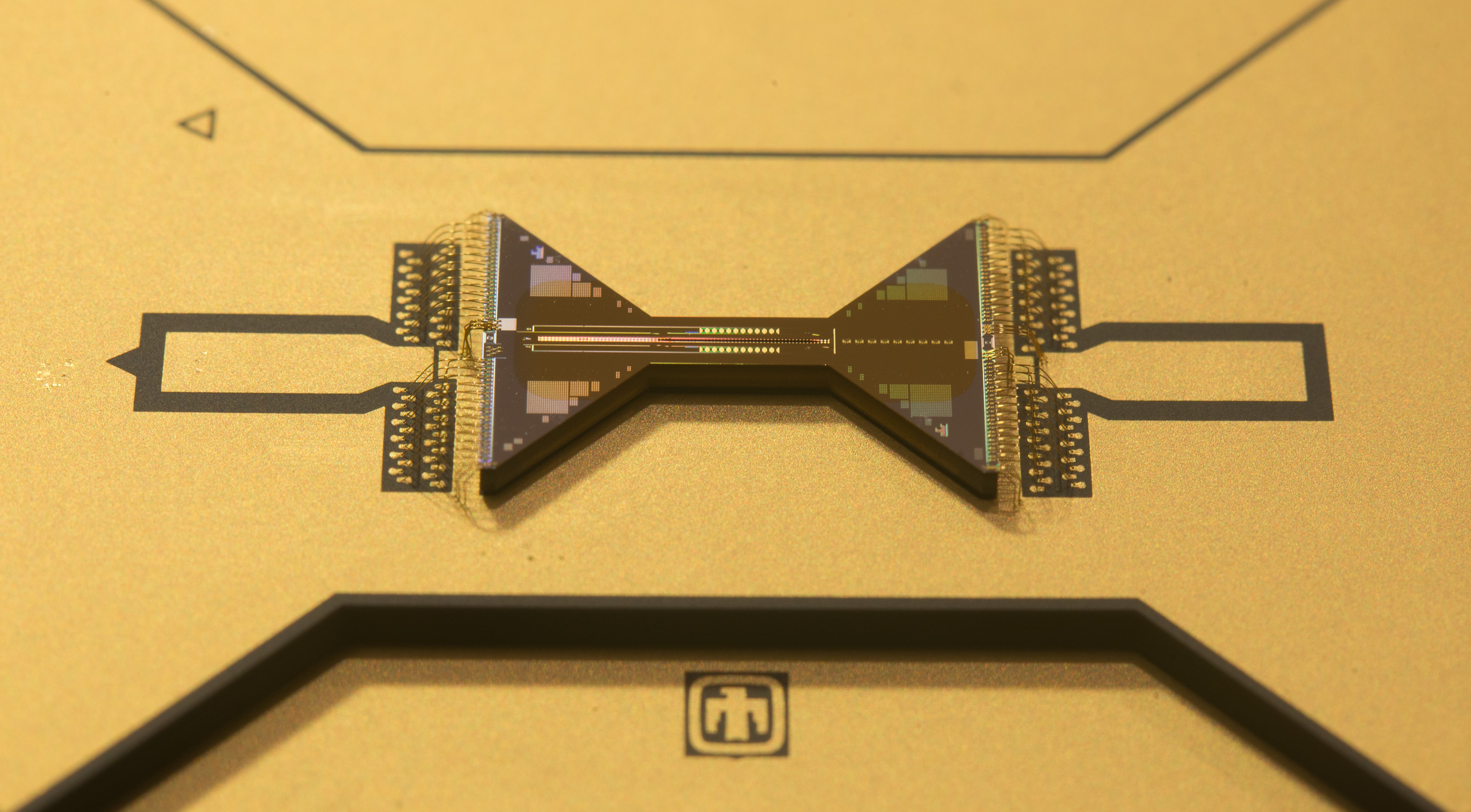}
	\caption{
		\label{fig:phoenix_optical}
		\textbf{Optical image of mounted Phoenix trap.}}
\end{figure}

\subsubsection{Trap Geometry}

\begin{figure}
	\centering
	\includegraphics[width=0.8\textwidth]{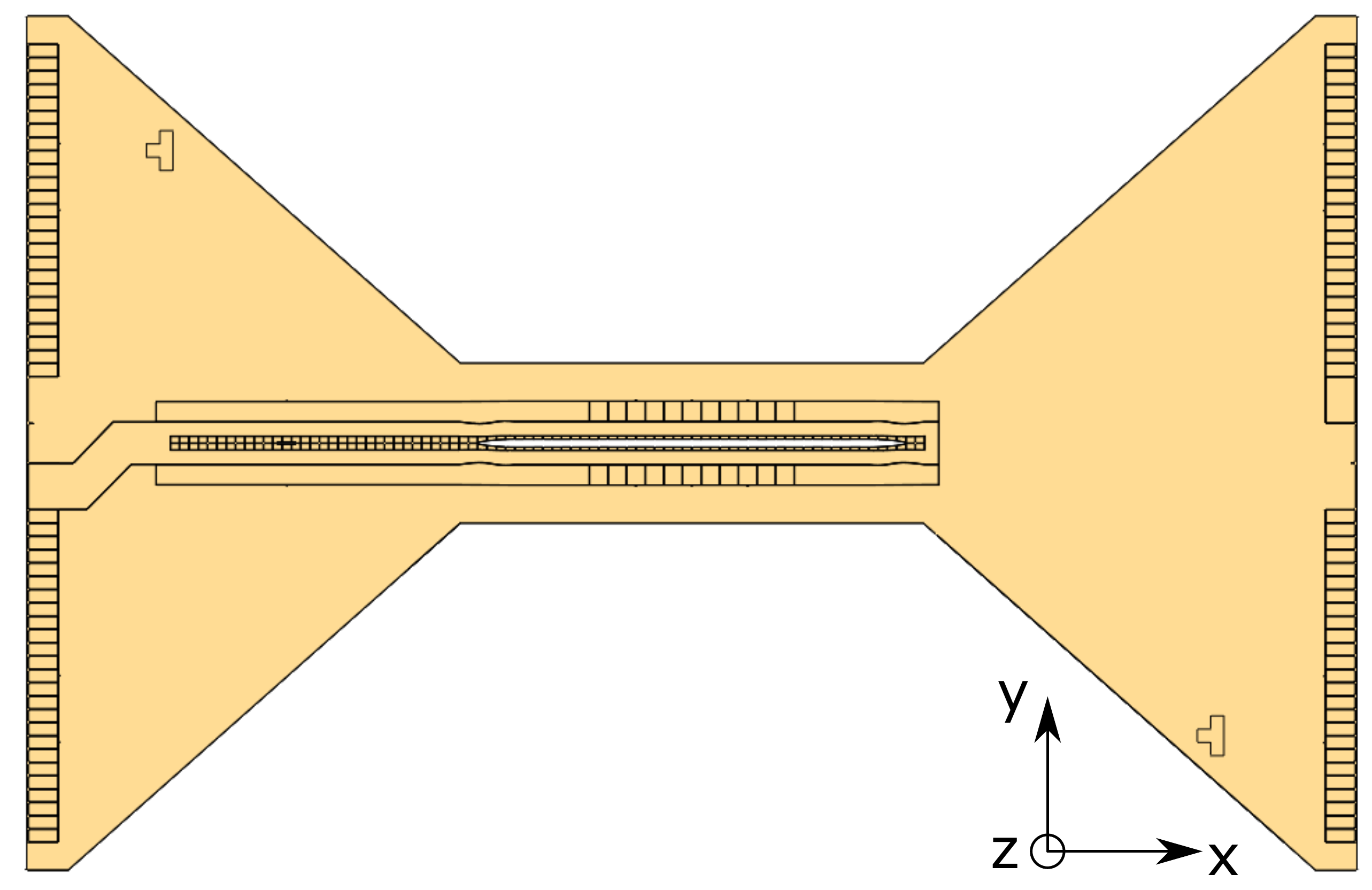}
	\caption{\label{fig:phoenix_rendering}
		Rendering of Phoenix trap fabricated on the HOA platform. The platform accommodates, $2$ rf connections that are used for rf feed and rf probe. 
		Ground wirebonds can be attached at two dedicated bondpads as well as the four corners of the substrate. 
		In addition, there are $100$ signal connections.
		The inner control electrodes are realized on a lower metal layer and are $\approx \unit[10]{\mu m}$ below the rf electrodes.
		The quantum region is centered on the slot with a transition to the above-surface region at each end. 
	}
\end{figure}

The Phoenix trap is a linear ion trap with a large slotted quantum region, a transition to an above-surface region, and an above-surface region (see Fig.~\ref{fig:phoenix_rendering}), each of these regions will be discussed in detail. 
The above-surface region includes a loading slot for backside loading.
The transition between slotted and above-surface regions is modulated to reduce axial rf pseudopotential barriers and to enable shuttling through the transition while keeping all trap frequencies and principal axes constant.

\begin{figure}
	\centering
	\includegraphics[width=\textwidth]{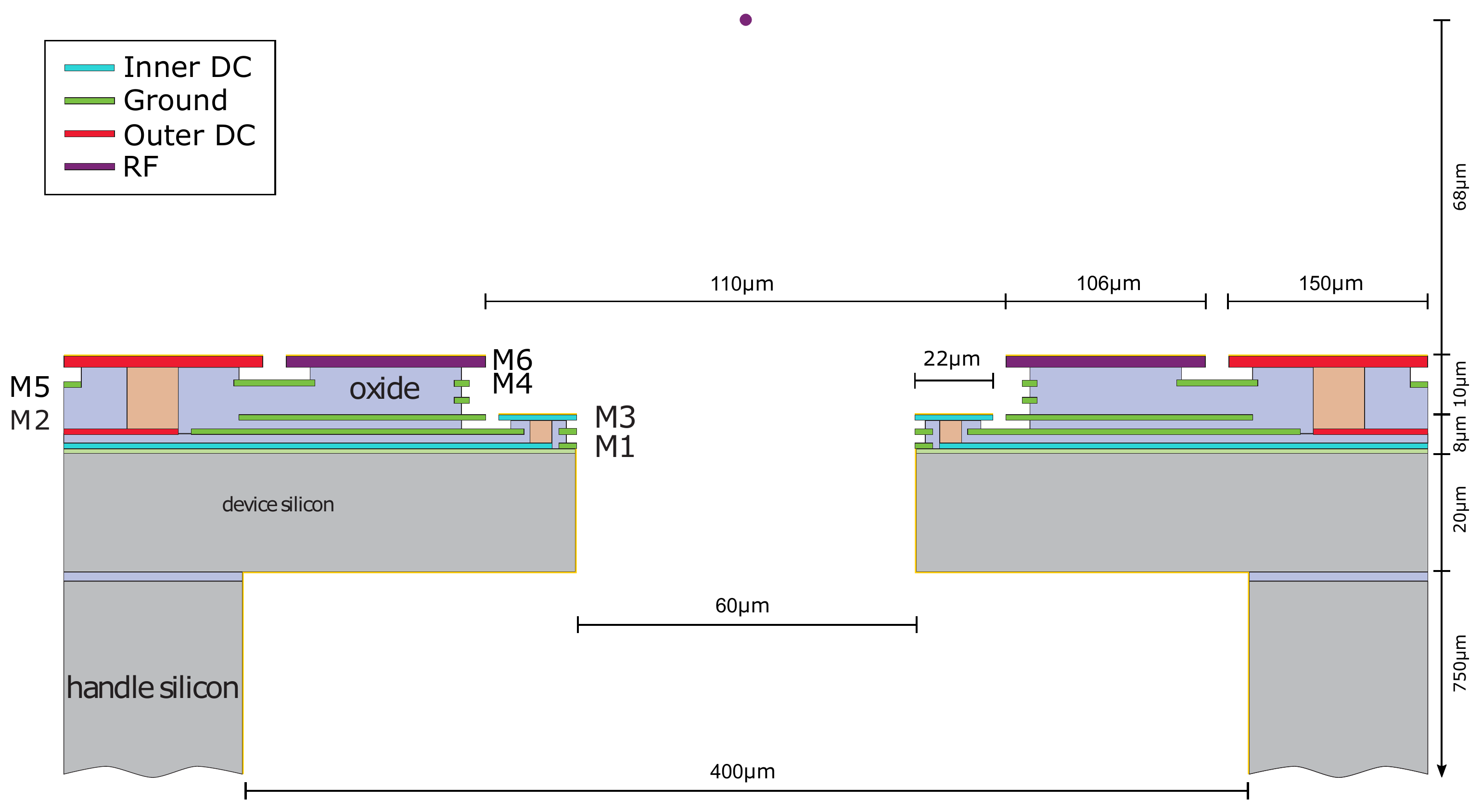}
	\caption{
		\label{fig:phoenix_profile}
		\textbf{Cross section of the Phoenix trap.} 
		(Not to scale.) The trap is fabricated using a 6-metal-layer process. 
		The rf, outer electrodes, loading region, and transition region are all located on the top metal layer, M6. 
		In the slotted region, the inner electrodes are inset at layer M3. 
		In this region, the ion is trapped about $68~\um$ above the top metal, M6. 
		The oxide thicknesses are nominal and vary slightly from trap-to-trap. }
\end{figure}

These devices were fabricated using 6 metal levels (Figure~\ref{fig:phoenix_profile}); the top (M6) is the electrode level (like the HOA-2, the Phoenix trap also uses the fourth metal level (M3) for the inner control electrodes), the lower metal layers (M1, M2 and M3) are
used for routing control lines. 
In locations where metal below the electrodes is exposed to the ion, the exposed metal is grounded. 
This applies for the central segmented control electrodes on M3 where M2 is grounded as well as for the
M3, M4, and M5 metal exposed through gaps in the M6 plane. 
All electrodes are overhung from the underlying silicon oxide insulating layers. 
Unless otherwise specified, the top metal is over-coated with 250 nm of gold, using titanium and platinum for adhesion.

\begin{figure}
	\centering
	\includegraphics[width=0.95\textwidth]{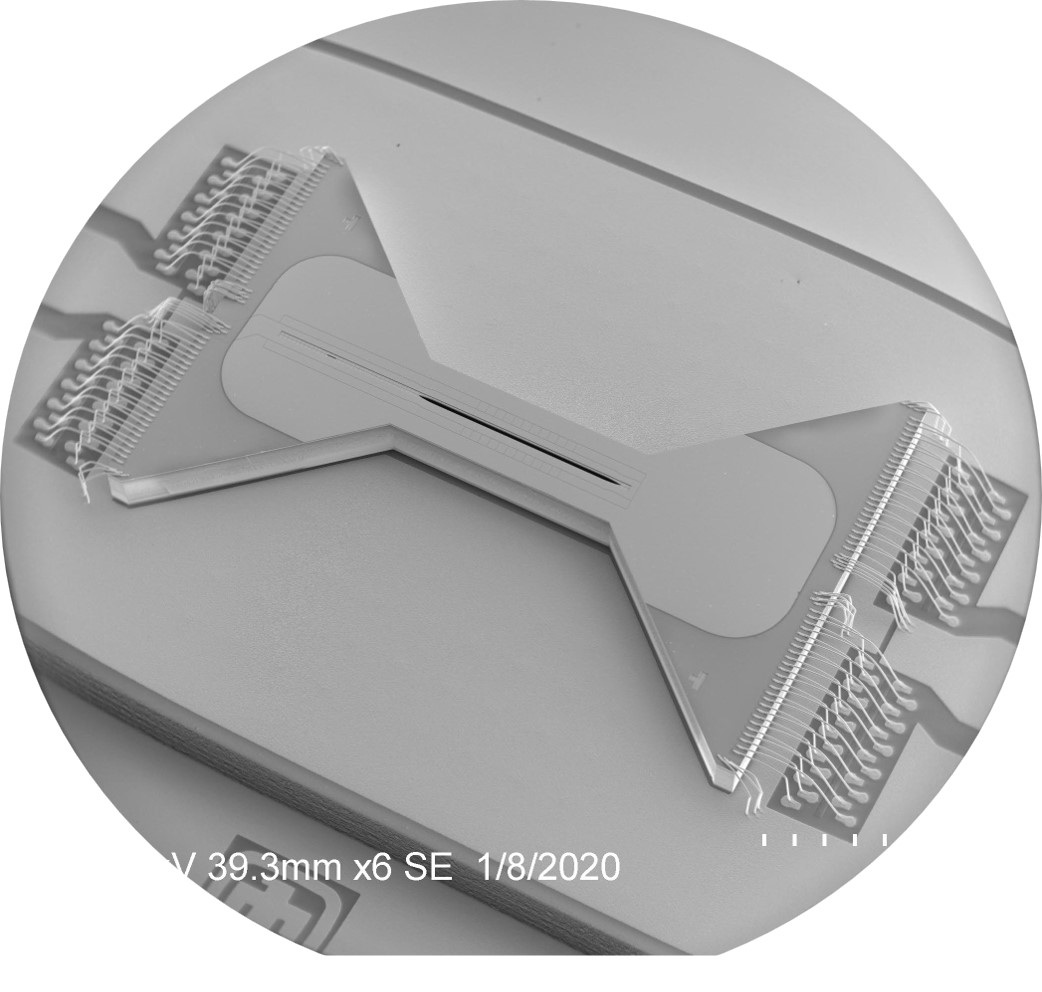}
		\caption{\label{fig:phoenix_SEM}Scanning Electron Micro-graph (SEM) of Phoenix Trap. The trap is mounted on the package and wirebonded.
		The trap center is kept clear of wirebonds to enable laser beams to be introduced along the linear axis of the trap. }
\end{figure}

\begin{figure}
	\centering
	\includegraphics[width=1.0\textwidth]{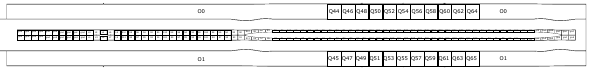}
	\caption{
		\label{fig:phoenix_schematic}
		\textbf{Schematic of Phoenix trap.} The slotted part of the trap is symmetric with the quantum region in the center.
		The loading region is on the left side with the hole centered on electrodes L04 and L05. The shuttling region connecting the loading and quantum regions uses
		$12$ electrode pairs that are repeated $5$ times.
	}
\end{figure}

We define the coordinate system with the $x$-axis in the plane of the top metal level along the long linear region of the trap, the $y$-axis in the plane of the trap surface perpendicular to the linear axis of the trap, and the $z$-axis perpendicular to the trap surface (see Figure~\ref{fig:phoenix_rendering}). The origin of the coordinate system is at the center of symmetry of the quantum region, between electrodes Q20, Q22 and Q21, Q23 (see Fig.~\ref{fig:phoenix_schematic}). In the vertical direction $z=0$ is defined as the top of the top metal level. 

\subsubsection{Quantum Region}
\begin{figure}[ht]
	\centering
	\includegraphics[width=0.8\textwidth]{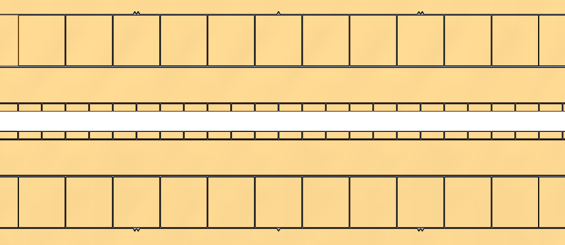}
		\caption{\label{fig:quantum_render}\textbf{Rendering of the quantum region:} It has $22$ pairs of inner electrodes and $11$ pairs of outer segmented electrodes.}
		    \vspace{.5cm}
	\includegraphics[width=0.8\textwidth]{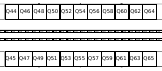}
		\caption{\label{fig:quantum_schematic}\textbf{Schematic of the quantum region:} beam alignment marks can be found between the outer control electrodes and the outer ground plane. Alignment marks are centered at 0~$\mu$m (single triangle) and $\pm$320~$\mu$m (double triangle).
	}
\end{figure}

The $\unit[1.54]{mm}$ long central quantum region is comprises $22$ inner control electrode pairs with independent control voltages and $11$ outer control electrode pairs (see Figs.~\ref{fig:quantum_render} and~\ref{fig:quantum_schematic} for the rendering and schematic, respectively, of the quantum region). The inner electrode pairs have a pitch of $\unit[70]{\um}$ and the outer pairs have a $\unit[140]{\um}$ pitch. The width of the central slot is $\unit[60]{\um}$. The inner control electrodes are located on M3, a lower metal level, thus they are $\approx -10\um$ vertically offset from the rf and outer electrodes.

The residual axial pseudopotential is simulated to be below $\unit[1]{\mu eV}$ within a range of $\unit[\pm1]{mm}$ of the trap center.
In the quantum region itself, it is simulated to be below $\unit[100]{neV}$.

The slot enables one to send laser beams vertically though the trap. 
The numerical aperture is limited by the backside etch of the trap and in the direction perpendicular to the trap axis has a numerical aperture of $0.25$ (see Sec.~\ref{subsec:opticalaccess} for further details).

\subsubsection{Transition Region}
\begin{figure}[ht]
	\centering
		\includegraphics[width=0.99\textwidth]{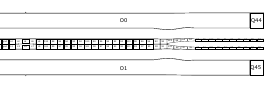}
			\caption{\label{fig:transition_schematic}
			\textbf{Schematic of the transition regions on the Phoenix trap.}
			The $6$ pairs of shuttling (S) electrodes repeat allowing for multiple trap minima to shuttle multiple individual ions between the loading region and the quantum region.}
		\includegraphics[width=0.45\textwidth]{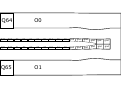}
			\caption{\label{fig:transition_schematic2}
			\textbf{Schematic of the end of the Phoenix trap.}
			The $6$ pairs of S electrodes repeat allowing ion(s) to be moved towards the end of the trap. 
			There is no surface region at the far end of the Phoenix trap, thus the electrodes in the far transition region are grounded.  }
\end{figure}
The linear section of the trap connecting the central quantum region to the loading region has many more electrodes than independent control voltages are available in the standard ceramic pin grid array (CPGA) or ceramic land grid array (CLGA) package. 
Therefore, it is necessary to use the same voltages on multiple electrodes. 
While the central quantum region uses $33$ control signal pairs, the shuttling regions on either side of the quantum region only use $6$ control electrode pairs that are repeated on either side (Figure~\ref{fig:transition_schematic}). 
Thus, multiple trap minima can be moved through the shuttling region simultaneously.

The availability of a slotted region is important for individual addressing of ions in a chain using the high numerical aperture imaging optic used to image trapped ions. 
However, to make a design with slotted regions compatible with increasing ion numbers, transitions between slotted and above-surface regions are necessary. 
A naive design of this transition would lead to large pseudo-potential bumps and would make it impossible to maintain trapping conditions while moving ions across this transition.

\begin{figure}
	\centering
	\includegraphics[width=0.45\textwidth]{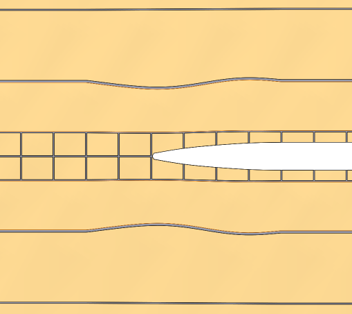}
	\includegraphics[width=0.535\textwidth]{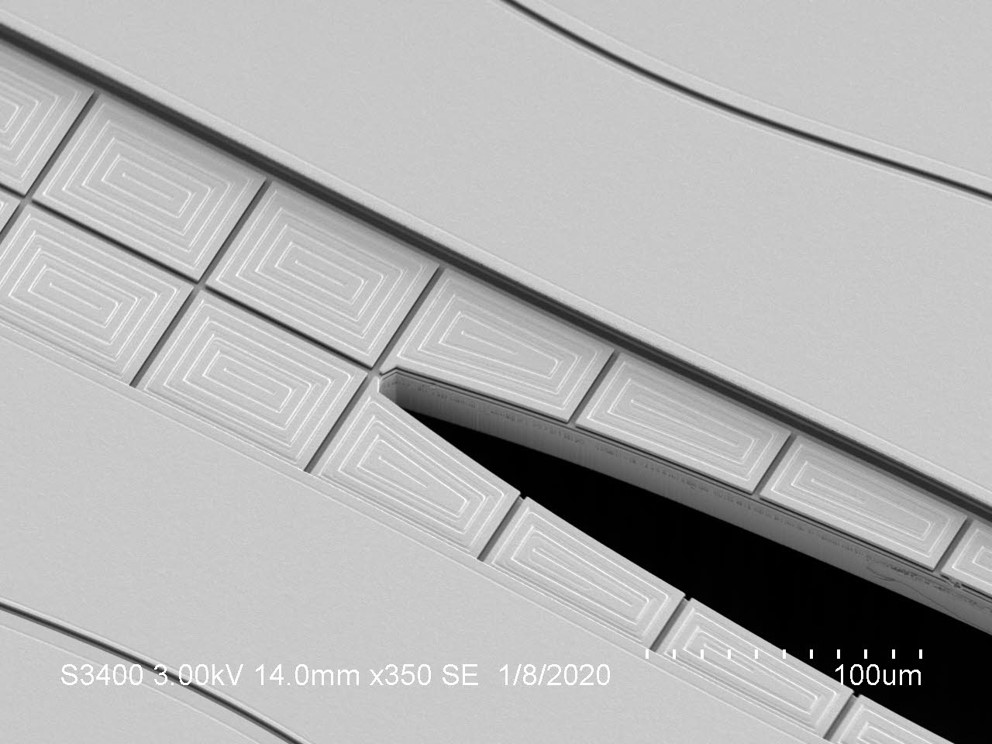}
	\caption{\label{fig:transition}
		\textbf{Transition region in the Phoenix trap.}
		The transition between the slotted region and the above-surface region of the trap is tapered and modulated to minimize disruptions of the linear ion trap. 
		The inner control electrodes are all realized $\unit[10]{\mu m}$ below the surface of the rf electrodes.
		Electrode modulations are defined using Fourier components to generate smooth modulations.
	}
\end{figure}

\begin{figure}
	\centering
	\includegraphics[width=0.5\textwidth]{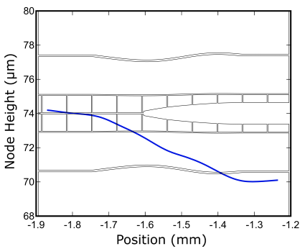}\hfill
	\includegraphics[width=0.5\textwidth]{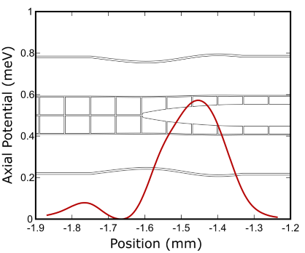}
	\caption{
		\label{fig:transition_potential_height}
		\textbf{Transition Properties:} 
		\emph{Left:} 
		Ion height above the top trap surface as a function of the position across the transition region.
	    The geometry of the transition is indicated. 
	    The ion height in the slotted region is $\unit[68]{\mu m}$ while it is $\unit[72]{\mu m}$ in the above-surface part of the trap. 
	    \emph{Right:}
	    Axial ponderomotive potential along the transition region calculated for ytterbium using $\unit[50]{MHz}$ rf signal at $\unit[250]{V}$ amplitude.
	    The concentric squares visible on the inner electrodes are the result of an over etch during processing. 
	    This leaves a small (less than 0.5~$\um$) feature on the inner dc electrode surface, this is small enough that we do not believe it effects trapping.
	    Subsequent versions of this trap will not have these features. 	}
\end{figure}

The transport region includes the transition between slotted and above-surface parts of the trap.
To enable shuttling of ions with minimal motional heating across this region, the inner control electrodes are tapered and the width of rf electrodes is modulated. 
The modulation is optimized to achieve a low residual axial ponderomotive potential and to be able to maintain constant trap frequencies while shuttling through the transition.
The trap frequencies and principal axes are controlled by the curvature tensor with $6$ degrees of freedom. 
As a consequence of the Poisson equation, the trace of the curvature tensor vanishes for purely static electric fields.
Instead, the trace of the curvature tensor, which needs to be non-vanishing to store an ion, is provided by rf fields. 
If rf voltage amplitudes are kept constant, the trace of the curvature tensor becomes solely a function of the geometry of rf and dc electrodes. 
The transition is optimized to keep variations of the trace of the curvature tensor across the transition region within a total variation of $\unit[3.4]{\%}$.
Residual rf pseudopotential barriers are below $\unit[3.5]{meV}$ (ytterbium, $\unit[50]{MHz}$, $\unit[250]{V}$). 
The optimized values for the axial pseudopotential barrier and the node height are shown in Figure~\ref{fig:transition_potential_height}.

\begin{figure}
	\centering
	\includegraphics[width=0.508\textwidth]{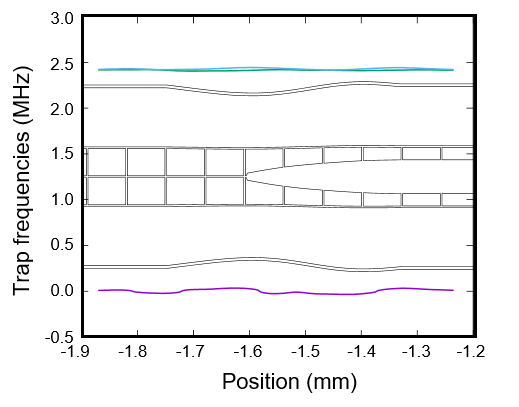}\hfill
	\includegraphics[width=0.485\textwidth]{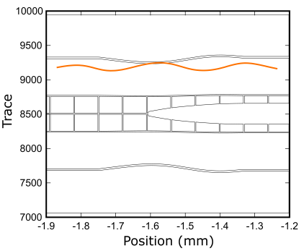}
	\caption{
		\label{fig:transition_trace}
		\textbf{Rf trap properties in the transition region:}
		\emph{Left:} 
		Axial and radial trap frequencies for ytterbium using $\unit[50]{MHz}$ rf signal at $\unit[250]{V}$ amplitude. 
		The variations in axial and radial trap frequencies across the transition are small and smooth.
		\emph{Right:}
		Trace of the curvature tensor as a function of the position within the transition region. 
		Variations of the trace are about 1\%.
		Consequently, it is possible to maintain all properties of a trapping potential as it is moved through the transition region with very good precision.
	}
\end{figure}

Diagonal elements of the curvature tensor, the trace of the curvature tensor, and trap frequencies are shown in Fig.~\ref{fig:transition_trace}. 

\subsubsection{Loading Region}

The loading region in the above surface part of the trap is designed for back side loading of an ion, while the slot at the center of the trap can also be used for loading, if it is convenient. The loading slot is centered $-\unit[3.045]{mm}$ from the center of the trap. The loading slot is hidden in the gap between the segmented inner electrodes. It has a length of $\unit[150]{\mu m}$ and a width of $\unit[13]{\mu m}$ at the surface metal. The loading slot is expanded in the handle silicon to a length of about $\unit[900]{\mu m}$ and a width of $\unit[100]{\mu m}$.
There are 5 pairs of independent electrodes centered on the loading region for trapping of an ion separate from the shuttling and transition region. 

\begin{figure}
	\centering
	\includegraphics[width=0.377\textwidth]{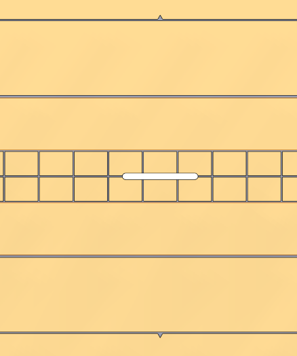}\hfill
	\includegraphics[width=0.6\textwidth]{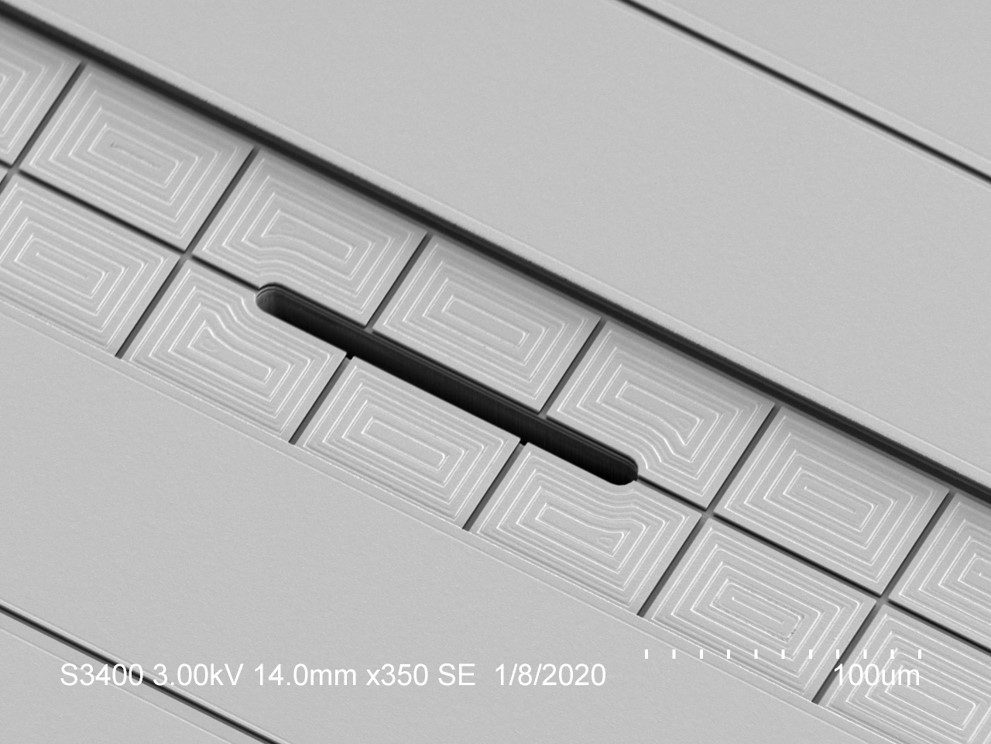}
	\caption{
		\label{fig:phoenix_loading}
		\textbf{Loading region in Phoenix trap:}
		\emph{Left:}
		Rendering of the loading region. The loading slot has a length of $\unit[140]{\mu m}$ and a width of 
		$\unit[13]{\mu m}$. $\unit[50]{\mu m}$ below the top surface of the trap, this profile is widened to a width of $\unit[100]{\mu m}$ and a length of $\unit[900]{\mu m}$.
		\emph{Right:}
		Scanning Electron Micro-graph of the loading region.
	}
\end{figure}

\begin{figure}
	\centering
	\includegraphics[width=0.6\textwidth]{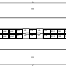}
	\caption{
		\label{fig:loading_schematic}
		\textbf{Schematic of the Phoenix Trap loading region.}
		A set of 5 pairs of independent electrodes can be used for creating an axial potential in the loading region. 
		After loading, the ion can then be shuttled to the quantum region and a new potential can be formed at the loading slot separate from the shuttling actions. 
	}
\end{figure}

\subsubsection{Rf Pseudo-Potential}
\label{sec:phxpseudoPotential}

The pseudo-potential strength of the trap was maximized by optimizing electrode configurations (rf electrode widths, gaps, realization of inner control electrodes on a lower metal layer) within fabrication constraints and maintaining a minimum ion height of $68~\um$. 
Figure~\ref{fig:phoenix_profile} shows a cross-section of the slotted region and the electrode widths and layers that were used to maximize trap strength. 
Compared to previous Sandia fabricated traps, the most important parameters for increasing the trap strength at a given rf voltage were increasing the rf electrode width and moving the inner dc electrode to the rf grounded M3 level.

The electrode widths were optimized to jointly give a high trap depth and high trap frequency. 
Optimizing for each individually would result in about a 10\% gain in either the trap depth or frequency for a given rf voltage. 
Since both are important for the applications of this trap, the small decrease in both, yields a compromise that results in each quantity being nearly 90\% of maximum.

\subsection{Peregrine Trap}
\begin{figure}[ht]
	\centering
	\includegraphics[width=\textwidth]{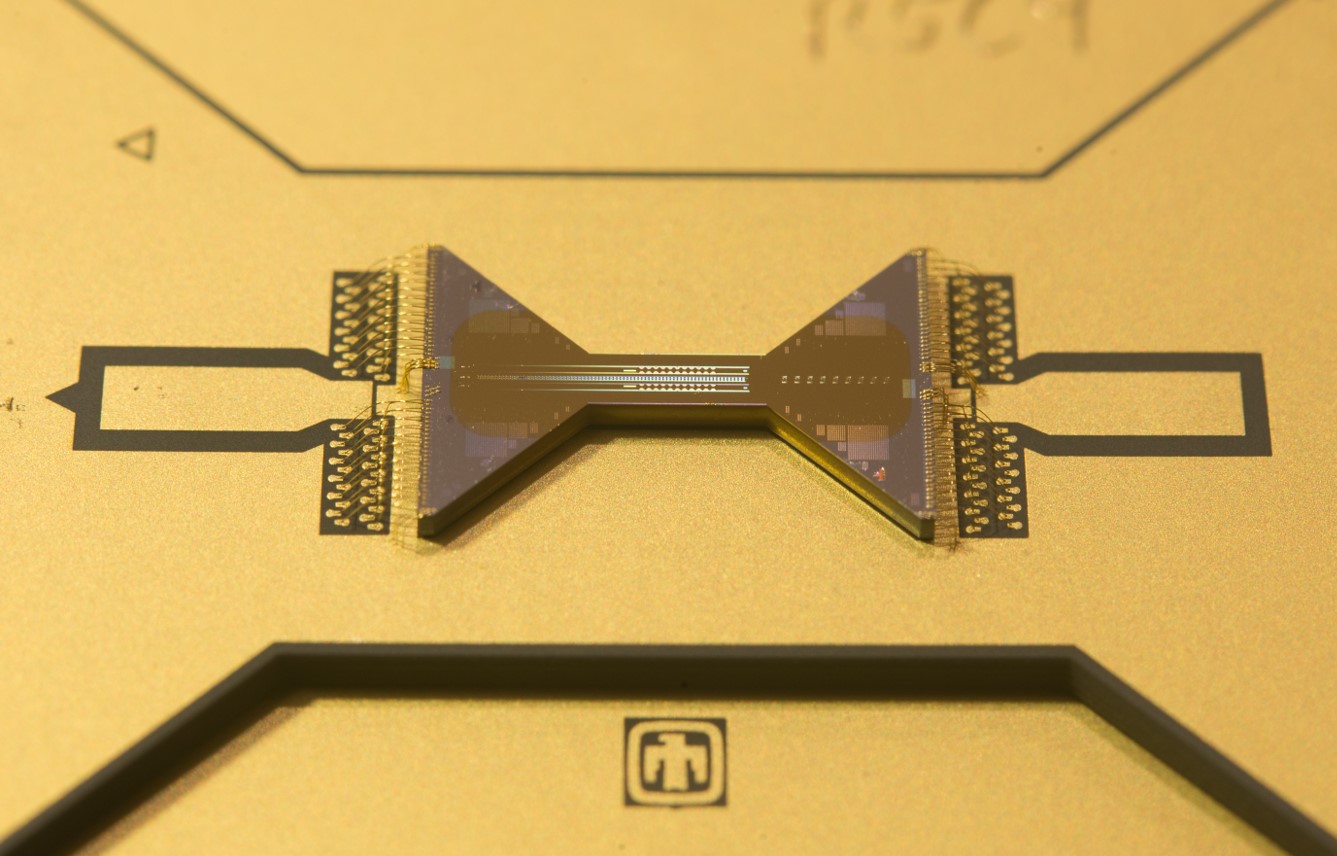}
	\caption{
		\label{fig:peregrine_optical}
		\textbf{Optical Image of mounted Peregrine trap.}
	}
\end{figure}

\begin{figure}
	\centering
	\includegraphics[width=0.8\textwidth]{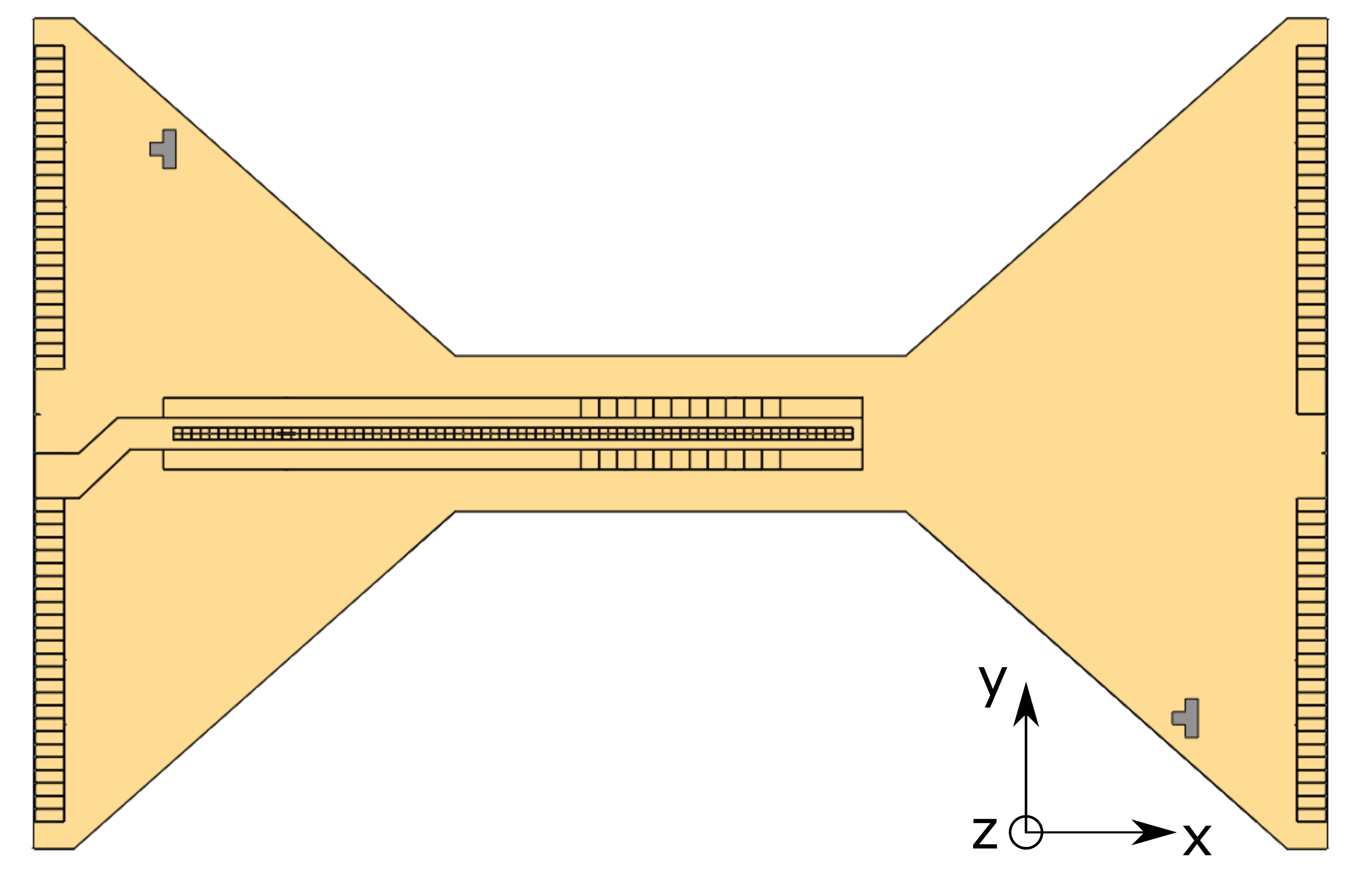}
	\caption{
		\label{fig:peregrine_rendering}
		\textbf{Rendering of Peregrine trap} fabricated on the High-Optical-Access platform. This is the same platform as the Phoenix trap. It accommodates $2$ rf connections (for rf feed and rf probe), ground wirebond attachments at two dedicated bondpads as well as the substrate corners, and $100$ signal connections.
		The Peregrine trap is a pure surface trap, so all the electrodes are realized on the top metal layer, M6. 
		The quantum region is centered on the isthmus of the trap with a shuttling region to the loading slot. 		
		}
\end{figure}

\begin{figure}
	\centering
	\includegraphics[width=\textwidth]{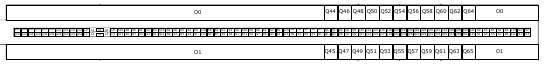}
	\caption{
		\label{fig:peregrine_schematic}
		\textbf{Schematic of Peregrine trap.} The quantum region is in the center of the device platform, it has $22$ pairs of inner control electrodes and $11$ pairs of outer electrodes. 
		The loading region is on the left side with the hole centered on electrodes L04 and L05. 
		The shuttling region connecting the loading and quantum regions uses $12$ electrode pairs that are repeated $5$ times.
	}
\end{figure}

\subsubsection{Trap Geometry}

The Peregrine trap is a 6-rail linear ion trap with a pure surface quantum region, a shuttling region, and a loading region (see Fig.~\ref{fig:peregrine_rendering} and~\ref{fig:peregrine_schematic}), each of these regions will be discussed in detail. 
The loading region includes a loading slot away from the quantum region for backside loading.
The shuttling region consists of pairs of electrodes that are connected for shuttling ions or chains back and forth from the loading region. 
This trap has the same electrode layout as the Phoenix trap, but is missing the loading slot. 
While slots can be useful for ion addressing and loading, they can also lead to higher potential for charging and the difficulty of shuttling through a transition region.
The Peregrine offers all the flexibility of the Phoenix trap, but without the slot for those who do not intend utilize its benefits.

These devices were fabricated using 6 metal levels; the top (M6) is the electrode level (unlike the Phoenix trap, all the electrodes are on the top metal layer), and the lower metal layers (M1, M2 and M3) are used for routing control lines. 
In locations where metal below the electrodes is exposed to the ion, the exposed metal is grounded. 
All electrodes are overhung from the underlying silicon oxide insulating layers. 
Unless otherwise specified, the top metal is over-coated with 50 nm of gold, using Ti and Pt for adhesion.

We define the coordinate system with the $x$-axis in the plane of the top metal level along the long linear region of the trap, the $y$-axis in the plane of the trap surface perpendicular to the linear axis of the trap, and the $z$-axis perpendicular to the trap surface (see Figure~\ref{fig:peregrine_rendering}). The origin of the coordinate system is at the center of trap isthmus, between electrodes Q20, Q22 and Q21, Q23. In the vertical direction $z=0$ is defined as the top of the top metal level.

\subsubsection{Loading Hole}

The loading region in is designed for back side loading of an ion and is the only hole through the handle silicon of the trap. 
The loading slot is centered $-\unit[3.045]{mm}$ from the center of the trap. 
Just like the Phoenix trap, the loading slot is hidden in the gap between the segmented inner electrodes. 
It has a length of $\unit[150]{\mu m}$ and a width of $\unit[13]{\mu m}$ at the surface metal. 
The loading slot is expanded in the handle silicon to a length of about $\unit[800]{\mu m}$ and a width of $\unit[100]{\mu m}$.
There are $5$ pairs of individually controllable inner electrodes centered on the loading region for trapping of an ion separate from the shuttling and transition region.

\begin{figure}
	\centering
	\includegraphics[width=0.5\textwidth]{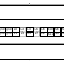}
	\caption{
		\label{fig:peregrine_loading}
		\textbf{Loading region of Peregrine trap.} The loading slot is at the center of the loading region is $\unit[150]{\mu m} \times \unit[13]{\mu m}$.  
		The loading region is at the far-left end of the trap. 
		There are $5$ pairs of individually controllable inner electrodes with a $70~\um$ pitch. 
		The outer control electrodes are rails (O0 and O1) that are shared with the shuttling region).
	}
\end{figure}

\begin{figure}
	\centering
	\includegraphics[width=0.39\textwidth]{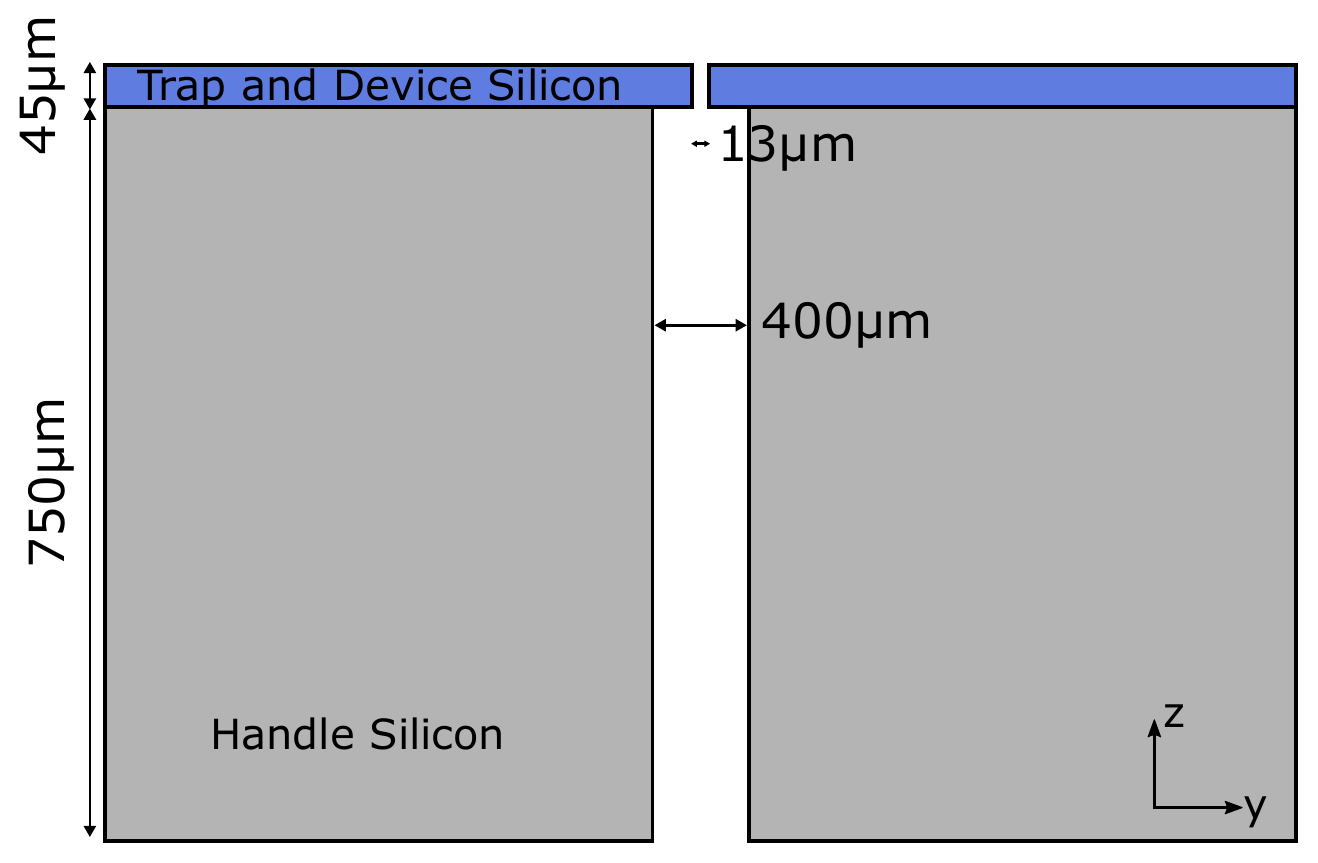}\hfill
	\includegraphics[width=0.54\textwidth]{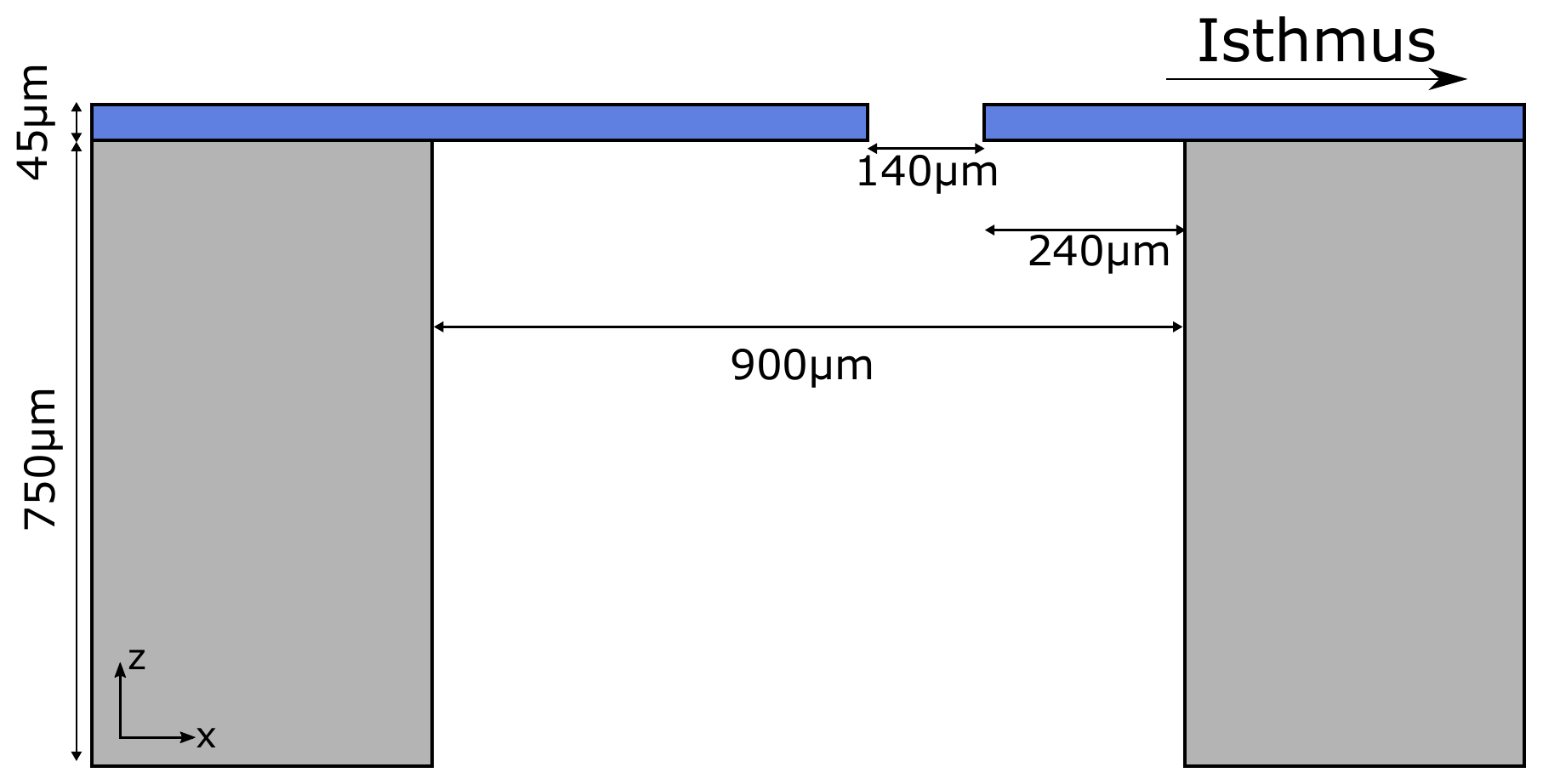}
	\caption{
		\label{fig:peregrine_loading_profile}
		\textbf{Loading region of Peregrine trap.} Profile through the center of the loading slot in lateral direction
		\emph{(left)} and axial direction \emph{(right)}. In the trap and device silicon, the slot dimensions are 
		$\unit[150]{\mu m} \times \unit[13]{\mu m}$, in the handle silicon they are 
		$\unit[900]{\mu m} \times \unit[100]{\mu m}$.
	}
\end{figure}

\subsubsection{Quantum Region}

The $\unit[1.54]{mm}$ long central quantum region is constitutes $22$ inner control electrode pairs with independent control voltages and $11$ outer control electrode pairs (see Fig.~\ref{fig:peregrine_quantum} for the schematic of the quantum region). 
The inner electrode pairs have a pitch of $\unit[70]{\um}$ and the outer pairs have a $\unit[140]{\um}$ pitch. 
The ion height is approximately $74~\um$ in this region. 
All the electrodes are located on the top metal as there is no slot in the quantum region. 

The residual axial pseudopotential is simulated to be below $\unit[1]{\mu eV}$ within a range of $\unit[\pm1]{mm}$ of the trap center. 
In the quantum region itself it is simulated to be below $\unit[100]{neV}$.

The HOA platform of the device allows high numerical aperture (NA) beams to access the quantum region from the side (see Sec.~\ref{subsec:opticalaccess}).
Vertically, the NA is limited by the width of the isthmus compared to the ion height. 

\begin{figure}
	\centering
	\includegraphics[width=\textwidth]{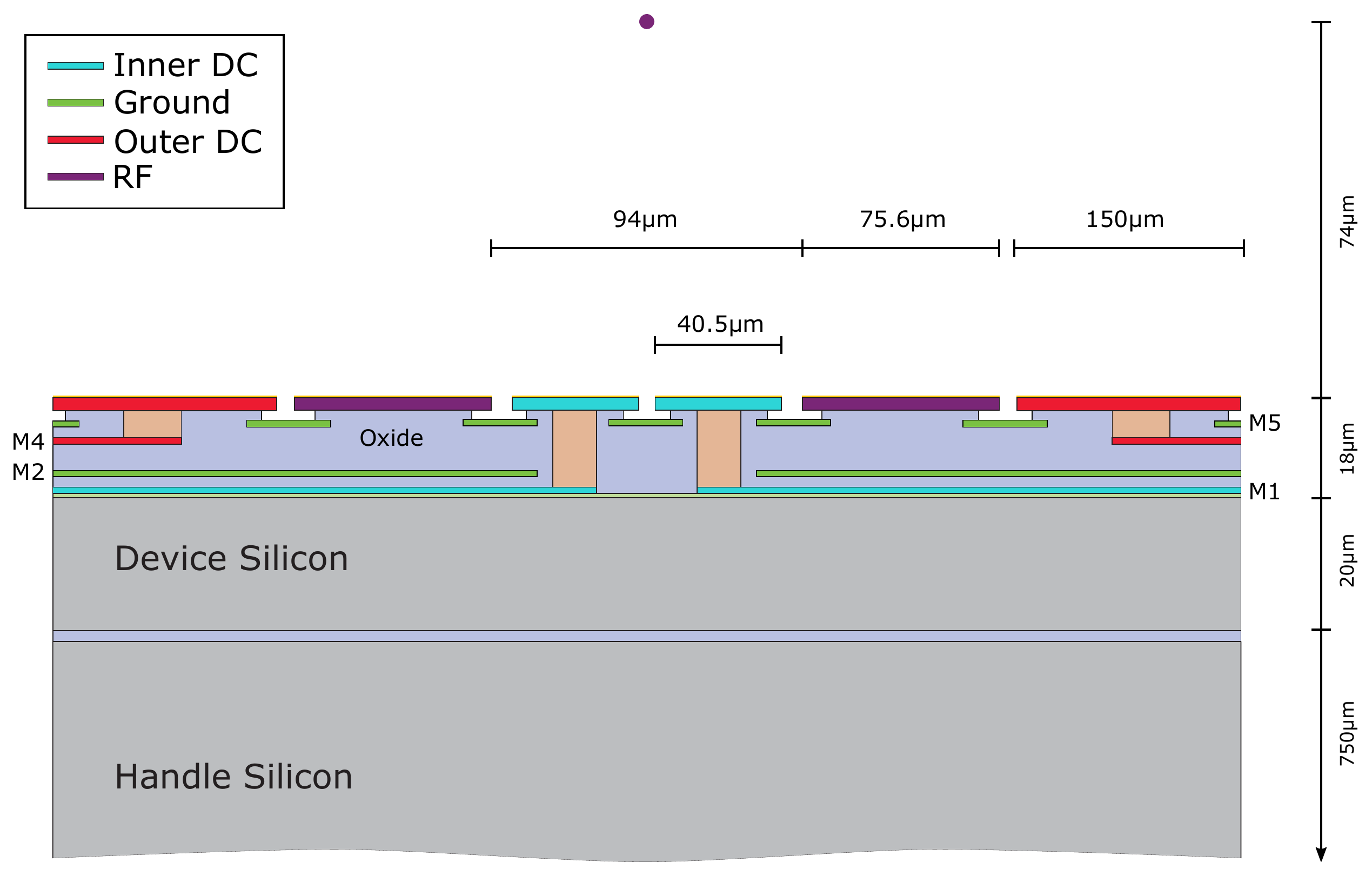}
	\caption{
		\label{fig:peregrine_profile}
		\textbf{Cross section of the Peregrine trap} in the quantum region (not to scale).  
		The trap is fabricated using a 6-metal-layer process. 
		All of the electrodes are located on the top metal layer, M6, and the ion is located about $74~\um$ above.
		There is no slot, so there is no exposed sidewall silicon.
		The M5 layer that is exposed through the gaps in the electrodes is grounded.
		}
\end{figure}

\begin{figure}[h]
	\centering
	\includegraphics[width=\textwidth]{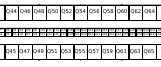}
	\caption{
		\label{fig:peregrine_quantum}
		\textbf{Quantum region of Peregrine trap} The Quantum region has $22$ inner segmented electrode pairs and $11$ outer segmented electrode pairs. The signals are Q0 \dots Q65 and can all be controlled independently.
		Alignment marks at the center of electrode Q54 (single) and Q48, Q60 (double) can be used to simplify beam alignment.
	}
\end{figure}

\subsubsection{Rf Pseudo-Potential}

The pseudo-potential of the Peregrine trap was optimized in the same manner as the Phoenix trap (see Section~\ref{sec:phxpseudoPotential}).
This resulted in a slightly different ion height and electrode widths. 
The primary factor for this difference is the lack of slot. 
Thus, the inner electrodes are on the same plane as the rf and give a slightly different optimization. 
However, the results of the Peregrine optimization are the same as the above surface region of the Phoenix trap. 
The ion height in these cases is about $74~\um$ as seen in Fig.~\ref{fig:transition_potential_height}.

\section{Control Solution Generation}
\label{sec:ContrlSolGen}

\subsection{Prerequisites}

To generate voltage solutions, a solved boundary model is needed. 
This is generally created from a meshing of the trap under consideration. 
The solved boundary element model is typically saved in a `.bec' custom format binary file.

\subsubsection{File Formats and Viewing}

The trap geometry is saved in a ``Double Polygon" file which is typically given the file extension `.dp'. 
In the next step, the trap geometry is meshed using the `Mesher'. 
The output format of the mesher is a boundary element file, which is given the extension `.be'. 
Both `.dp' and `.be' as well as `.bec' files can be converted to `.vtk' or `.msh' files using the python utilities `convert\_vtk.py' and `convert\_msh.py', respectively.

The `.vtk' files can be viewed using Paraview (https://www.paraview.org/), while the `.msh' files can be viewed using Gmsh (https://gmsh.info/). 
In both cases, only the (meshed) geometry of the files is converted, not the boundary element solution.

\subsection{Approach}

From the boundary element solution (`.bec' file), a ``voltage cube" is calculated. 
The voltage cube contains the potential for one signal at $1$~V while all other signals are grounded and is defined on a 3-dimensional rectangular grid.
The voltage cube can either be pre-calculated (typically with `.va' as the file extension), or calculated on the fly from the `.bec' file using a Python module.

From the voltage cube data, the ponderomotive potential can be calculated from the gradient of the voltage and scaled by the RF amplitude and drive frequency.
Traditionally, Mathematica has been used to calculate a 3-dimensional spline to interpolate between the points on the grid. 
However, due to numerical problems this does \emph{not} lead to a smooth potential surface, but a surface with small minima. 
This becomes a problem when searching for the nodal point or line: a local optimization algorithm can and will get stuck in one of these artificial minima.

The approach, taken here, to get around this problem is to approximate the 3-D potential on a voltage cube using a 3-D polynomial with a fixed number of degrees of freedom. 
Based on the polynomials, the calculation of ponderomotive potential is straightforward and the minimum of the ponderomotive potential is well-defined and easy to calculate numerically.

\subsection{Ways to Solve for Solutions}

The shape of an ion trap is, to second order, described by the curvature tensor. 
This is the Hessian of the pseudo-potential combined with the electrostatic potential generated by the control electrodes. 
The curvature tensor is symmetric and thus has 6 degrees of freedom. 
In addition, there is the residual electric field at the position of the rf node, which contains 3 additional degrees of freedom.

Due to the linear nature of the problem, the curvature tensor $H$ can be written as the sum of the curvature tensor generated by the rf pseudo-potential and the curvature tensor generated by the dc control fields

$$ H = H_{rf} + H_{dc}.$$

Here $H_{dc}$ is a traceless symmetric tensor with 5 degrees of freedom. 
In an ideal linear trap, and written in the basis of the principal axes, $H_{rf}$ is diagonal with a trace $t$ given by the rf voltage. 
$H_{rf} = diag(0, \nicefrac{t}{2}, \nicefrac{t}{2})$.

\subsubsection{Pseudo-Inverse Solution}
If we are solving for a single isolated trap location, a well-designed trap will have electrodes of sufficient number and adequate shape available to control all degrees of freedom of the curvature tensor. 
In fact, usually the problem is under-determined. 
That means we have more electrodes available than needed to achieve the prescribed value.

If there were precisely as many electrodes as degrees of freedom, we would have a problem where a simple matrix inversion would lead to the unique solution. 
In the under-determined case, we can use the pseudo-inverse. 
The pseudo-inverse is the unique solution to the problem that has the minimal 2-norm of the solution vector. 
For voltage solutions, this means the pseudo-inverse provides a unique solution with the minimal 2-norm of the voltage vector.

Using the pseudo-inverse has several advantages:
\begin{itemize}
	\item It provides a unique solution and thus is independent of starting values.
	\item Because it is the solution with the minimal 2-norm of the voltage vector, it will only use nearby electrodes.
	\item When a solution is calculated for a nearby position or similar potential, the voltage solution will also be similar. This helps in generating smooth shuttling solutions.
	\item This approach can also be used to simultaneously solve for multiple trapping sites, provided they are sufficiently separated to allow for independent control of all degrees of freedom.
\end{itemize}

But it also has some challenges:
\begin{itemize}
	\item Minimizing the 2-norm of the voltage vector has the tendency to create trap potentials with the intended trap frequency, but a small trap depth. A way around this could be to define offset voltages that prevent this scenario.
	\item If we solve for multiple nearby potentials (for example for separation and merging of ion chains) there will likely not be a sufficient number of independent degrees of freedom and voltages will diverge.
\end{itemize}

\subsubsection{Nonlinear Optimization}
In the case of nearby trapping potentials as required for the separation and merging of ion chains, there are typically not enough degrees of freedom to fully control all degrees of freedom of both traps. 
Therefore, the voltage solutions generated with the pseudo-inverse will diverge. 
In other words, within the voltage budget, there is no solution controlling all degrees of freedom.

In this case, the preferred approach is the minimization of a cost function while giving different degrees of freedom different weights. 
For example, when separating or merging two ions, the trap frequency becomes a property of the control voltage geometry ---mainly because at the quartic point, the geometry of the quartic, which is defined by the electrode geometry and ion distance determines the trap frequency--- and can hardly be adjusted by control voltages. 
In a surface trap, the principal axes of the trap close to the linear direction of the trap will tilt. This tilt is also hard to control.

Thus, a high weight will be chosen for the trap location, a low weight is to be chosen for the degrees of freedom that cannot be controlled (trap frequency and principal axes direction), and a medium weight can be chosen for the remaining degrees of freedom.

\section{Trap Features}

\subsection{Optical Access}
\label{subsec:opticalaccess}

The High Optical Access (HOA) trap platform was designed to accommodate tightly focused laser beams across the surface or through the central slot (if applicable) of the trap in order to achieve high laser intensities at the ion positions and to enable individual addressing of ions in a chain.
Sandia first introduced this in the so named HOA trap~\cite{HOAManual}, and this is discussed in detail in the HOA manual.

The optical access is parameterized by the fraction of a focused Gaussian beam which clips on a portion of the trap. 
In the case of the Phoenix/Peregrine traps, both the trap and the package impose similar restrictions on the solid angle across the surface of the trap.
The vertical numerical aperture (NA), for a beam passing parallel to the trap surface is limited by the isthmus of the trap. 
The ion sits in the rf node either 68~$\mu$m or 74~$\mu$m above the surface, for the Phoenix and Peregrine respectively.
The width of the isthmus of the trap (the width of the bowtie at its narrowest) exceeds 1.2~mm, just as in the HOA trap. 
Thus, NA for a beam propagating perpendicular to the isthmus of the trap is 0.11, and 0.08 for a beam propagating at 45$^\circ$, this will accommodate beam waists of $<2.5~\mu$m at the ion. 
The package was optimized to not further impede 370~nm light perpendicular to the isthmus.

In addition to being concerned about achieving high optical access for a side laser, we also needed to preserve high optical access through the slot for the Phoenix trap (not applicable on the Peregrine). 
Figure~\ref{fig:phoenix_profile} shows the measurements for beam clearance analysis in the slotted region of the Phoenix trap.
For a beam perpendicular to the chip, the available numerical aperture is 0.25 and limited by the width of the slot in the surface metal (60~$\mu$m).
The sidewalls of the slot are gold coated to help prevent charging from passing a beam through this region. 

\subsection{Rf Probe}

The stability of the rf voltage on the device is paramount for creating stable radial modes. 
While the rf reflected from the device or transmitted to the device are a reasonable proxy for the voltage at the trap, there can be sources of variation that these do not account for. 

Additionally, the current methods for accurately determining the voltage applied to the trap is to measure the radial frequencies of the ion and compare to the trap model.
However, if an ion has yet to be trapped in the device, the voltage is estimated based on the previous devices or estimated losses in the system. 
Inaccuracies in the estimate can lead to far exceeding the desired voltage or trying to trap with too low of an amplitude. 
Knowing the applied voltage more precisely in advance of trapping an ion can reduce the uncertainty in trying to trap.

\begin{figure}
\centering
\includegraphics[width=0.8\textwidth]{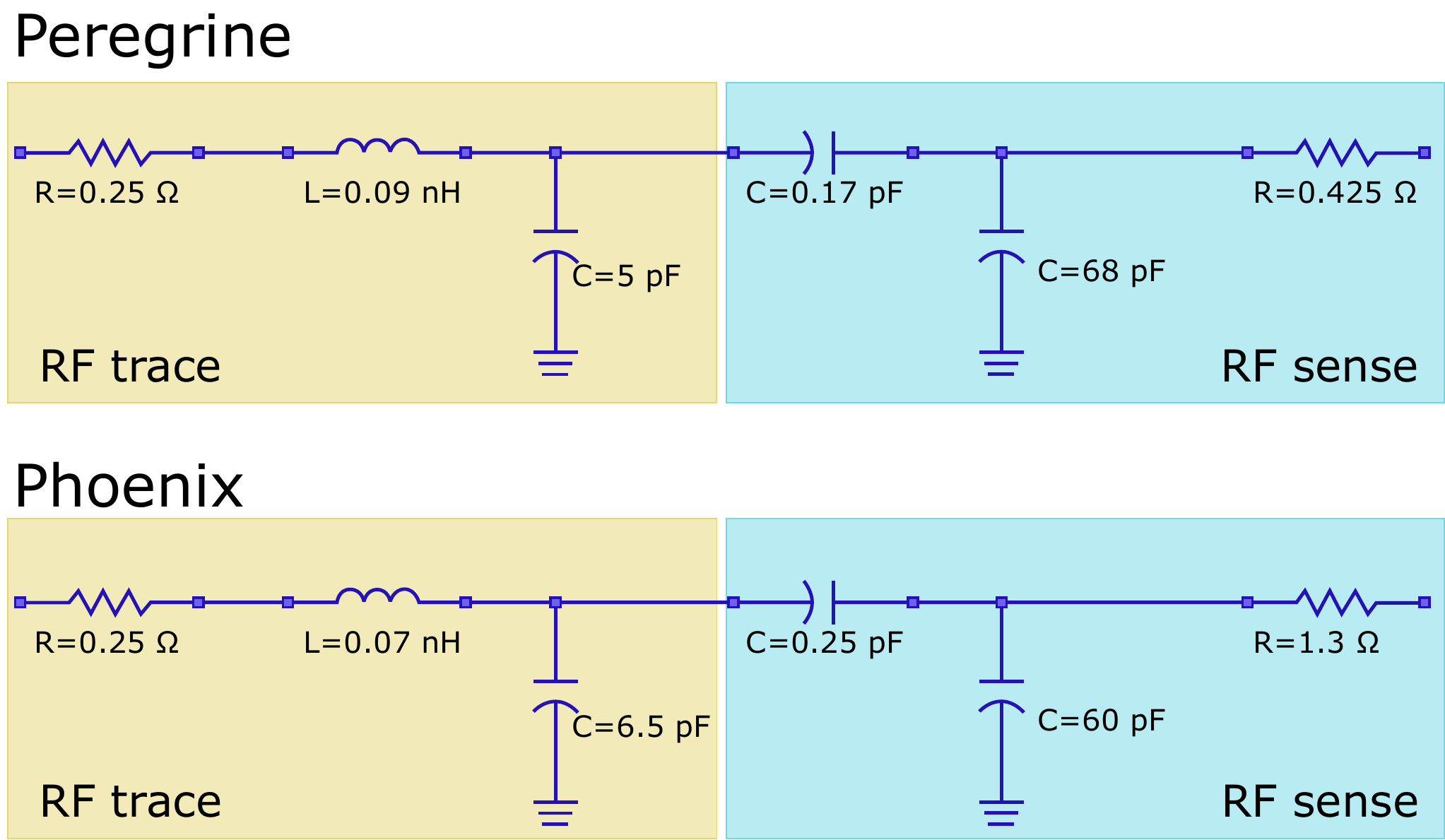}
\caption{\label{fig:rfsense}
This shows the equivalent circuit diagram for the rf probe on the Phoenix and Peregrine traps.
The rf sense is coupled via a capacitive divider (C=0.17~pF) to the rf trace. 
Though the traps have a very similar layout, there are slight differences between the properties of the rf trace. 
These differences are primarily due to the presence of the slot in the Phoenix trap. 
The slight changes in rf width from the slot lead to small differences in the effective capacitance and resistance.
}
\end{figure}

In the Phoenix/Peregrine trap, a rf sense wire was implemented to allow measurement of the rf voltage on the device. 
The rf probe is coupled to the rf trace via an approximately 200:1 capacitive divider. 
The equivalent circuit diagram is shown in Fig.~\ref{fig:rfsense}.

Though designed to be a 200:1 voltage divider, the measured result differed slightly. 
We measure an S21 on the Peregrine to be -55~dB at 35~MHz and -52~dB at 50~MHz, while it is -51~dB and -52.5~dB on the Phoenix, for the same frequencies respectively.
In a secondary measurement using a 50~$\Omega$ load measured by an oscilloscope, we measured -52~dB on the Peregrine at 50~MHz and -48~dB on the Phoenix.
While there are some differences device to device, this probe, once calibrated, is an accurate indication of the rf voltage on the device and can be used as part of a feedback circuit to stabilize the rf voltage on the device.

\subsection{Resistive Wires for Heating or Temperature Sensing}
\label{sec:tempwire}

Cryostats have a limited heat load capacity and while we can calculate the expected heat load of a trap, this may vary from device to device. 
Knowing the temperature of the device during operation may help tune the rf and to optimize heat dissipation through the cryostat. 
Additionally, keeping the trap at a higher temperature than its surrounding while cooling the cryostat may prevent condensation on the trap surface. 

Towards that end, we have incorporated heater and temperature sensing wires into the trap. 
While the benefits of these wire are clear in a cryogenic setting, they can also be used in room temperature. 

There are two sets of two wires in the trap, one set in aluminum and the other made from tungsten. 
In Table~\ref{tab:ElectrodeID}, the aluminum wires are listed as ``sense'' and ``resistive'' and both are located on the M5 metal layer (see Figs.~\ref{fig:phoenix_profile} and~\ref{fig:peregrine_profile}). 

The ``sense'' wire has a lower voltage limit, this is connected to trench capacitors in the device and thus is limited to voltages of $<20$~V, same as the electrodes. 
However, the trench capacitors will limit the rf pick-up on the wire, thus making it ideal for measuring the temperature. 
The entire wire should have a resistance of about 1.75~k$\Omega$.
Due to slight variations from device to device in the metal and dielectric thicknesses, giving an accurate equation for the calibration of this wire is not possible.
However, this can be calibrated for a specific device. 
Without applying rf to the trap, we mapped the temperature of the cryostat while cooling to the resistance of the wire. 
At high temperatures (room temperature), the resistance vs temperature response is nearly linear (very approximately 1~k$\Omega/$K) and as it approached 4~K, the resistance begins to flatten out around 80~$\Omega$.

The ``resistive'' wire is not connected to the integrated trench capacitors and thus can be operated at a higher voltage.
Since it will be more susceptible to rf pick-up, this wire is more suited for heating the device. 
However, the aluminum wire is limited by electromigration, which for aluminum begins around 2~\nicefrac{mA}{$\um^{2}$}.
Thus, the maximum current that can be applied to the wire is 2.4~mA.

The aluminum wires are added in locations that were not used on the previous packaging standards; thus, it is possible that some of the routing may need to be altered (or added) to accommodate these wires. 
Alternatively, there are the two tungsten wires which are labeled W1 and W2 on Tab.~\ref{tab:ElectrodeID} located beneath all the metal layers of the trap. 
These are connected to bond pads which have been used historically and are likely wired out of the chamber. 
Neither of the tungsten wires are connected to trench capacitors, so they can be used interchangeably as heaters or sensors. 
The room temperature resistance of the wires is approximately 2.6~k$\Omega$ and should decrease at lower temperatures. 
Additionally, tungsten does not suffer from the same electromigration limit and should be limited by the wirebond and dielectric breakdown. 
For these wires, the voltage and current limits are 100~V (either dc or rf) and 0.4~A.
The higher current limit of the tungsten allows for greater heating capacity and is recommended for use as the primary heater (should one be used).

\subsection{Rf Dissipation}

\begin{figure}
\centering
\includegraphics[width=0.5\textwidth]{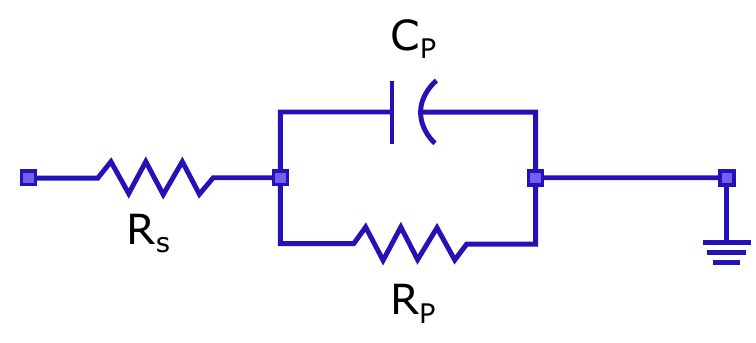}
\caption{\label{rccircuit} Schematic of the approximate rf circuit diagram. 
$R_{s}$ is the series resistance, resistance from the rf lead.
$R_{p}$ is the parallel resistance, loss from dielectric absorption.
$C_{p}$ is the parallel capacitance, the capacitance from the rf trace.}
\end{figure}

Having a large rf dissipation on the trap can lead to undesired heating of the device.
The long rf trace in the previous trap had the consequence of a large heat load being dissipated (about 100mW at room temperature). 
Both the resistance and the capacitance of the Phoenix and Peregrine have been reduced as much as possible to minimize dissipative loss. 
To estimate this, we can model the rf of the trap by assuming a serial resistance and a parallel resistance and capacitance (see Fig.~\ref{rccircuit}).
The power dissipated in the trap at a particular rf drive frequency ($\omega$) and can be approximated from:
\begin{equation}
P_{s} = \frac{1}{2}R_{s}U^{2}\omega^{2}C_{p}^{2} \propto L^{3},~ P_{p}=\frac{1}{2}\frac{\omega U^{2}}{R_{p}} << P_{s}.
\end{equation}
The resistive loss of the rf dominates over the absorptive loss from the dielectric.
To minimize this loss, the rf feed can be parallelized and the length of the rf ($L$) kept short. 

The measured resistances and capacitances are shown in Table~\ref{tab:rcchart} along with the estimated power dissipation compared to the HOA-2.1 trap.
The decrease in the length of the rf electrode is the primary change leading to the 3 times lowers dissipation in the Phoenix and Peregrine traps. 
The width of the rf was optimized based on the ion height. 
However, the length of the trap was chosen to minimize the rf length while keeping a large quantum region. 
The rf does not extend into the other bowtie because there are no junctions in this trap, so only the central region would be used. 
The rf feed on these traps is symmetric and from a single end minimizing the length and consequently, the capacitance. 
The lack of slot in the Peregrine trap leads to slightly different rf electrode width and thus a lower capacitance. 
This results in the Peregrine having 5 times less power dissipation than the HOA-2.1 device.

\begin{table}
	\caption{\label{tab:rcchart} 
	List of the measured capacitance and resistance of the rf trace on the trap (package not included in measurement). 
	The dissipated power is calculated from these measurements. 
	The power loss is lower in the Phoenix and Peregrine due, primarily, to the decrease in the electrode lead and the parallelized feed.  }
		\centering
		\begin{tabular}{|L{0.2\textwidth}|C{0.15\textwidth}|C{0.15\textwidth}|C{0.15\textwidth}|}
			\hline\xrowht{18pt}
			\Large{\textbf{Trap}} & \Large{$C_{p}$} & \Large{$R_{s}$} & \Large{$P_{s}$}\\ 
			\specialrule{.15em}{.2em}{.2em}\xrowht{11pt}
			Peregrine & 5.4~pF &0.36~$\Omega$ & 20~mW \\
			\hline\xrowht{11pt}
			Phoenix & 7.1~pF &0.4~$\Omega$ & 35~mW \\
			\hline\xrowht{11pt}
			HOA-2.1 & 7.6~pF &0.9~$\Omega$ & 100~mW \\
			\hline
		\end{tabular}
\end{table}

\section{Trap Fabrication}

The delivered devices contain the following materials:

\paragraph{Device} Silicon, silicon dioxide, silicon nitride, aluminum, copper, titanium nitride, titanium, platinum, tungsten, gold.

\paragraph{Die attach} Au80Sn20 solder, EpoTek H21D, or SAC solder (as specified in the device specific documentation -- PowerPoint presentation sent with the device).

\paragraph{Package} Aluminum Nitride, Gold, Nickel, Tungsten. Pins: iron, nickle, and cobalt alloy and silver-copper braze. 

\subsection{Trap Chip}
These devices were fabricated using 6 metal levels (Figure~\ref{fig:phoenix_profile} and~\ref{fig:peregrine_profile}); the top (M6) is the electrode level (although the Phoenix trap also uses M3 for the inner control electrodes), the lower metal layers (M1, M2, M3, and M4) are used for routing control lines.  
In locations where metal below the electrodes is exposed to the ion, the exposed metal is grounded. 
This applies for the central segmented control electrodes on M3 where M2 is grounded on the Phoenix trap as well as for the M5 metal exposed through gaps in the M6 plane on both traps.  
AlCu $(99.5\%/0.5\%)$ is used for the metal levels, with tungsten vias for vertical interconnections.  
All electrodes are overhung from the underlying silicon oxide insulating layers.  
Unless otherwise specified, the top metal is over-coated with $250\unit{nm}$ of gold, using 100~nm each of titanium and platinum for adhesion.

\subsection{Package and Die Attach}

These devices are packaged on top of a high-temperature co-fired ceramics (HTCC) Aluminum Nitride custom bowtie shaped package integrated onto either a CPGA or CLGA (see Figure~\ref{fig:CPGA_package_rendering} or~\ref{fig:CLGA_package_rendering}).
The trap is attached on top of the package using either solder or epoxy, (Au80Sn20 or SAC305 solder or EpoTek H21D~(H21D.pdf).

The Au80Sn20 solder is jetted onto solder pads on the ceramic package using a solder jetter. 
The bottom of the die has matching solder pads and the die is placed and the solder is reflowed in nitrogen or formic acid using the Finetech Femto 2 die bonder.

If the die was epoxy attached, the attach is cured at $225^{\circ}C$ for a duration of two hours to ensure a complete cure.  

We recommend a maximum baking temperature of $200^{\circ}C$ for less than 7 days to achieve ultra-high vacuum. 
This maximum temperature and time mitigate the formation of brittle and high resistivity gold-aluminum inter-metallic at the wire bonds.  
We recommend staying below $225^{\circ}C$ for the vacuum bake because that is the maximum temperature we've exposed it to, but its glass transition temperature is $240^{\circ}C$ and its glass melting point is $370^{\circ}C$, so in principal it could go much higher. 
However, higher temperatures risk accelerating the formation of a gold-aluminum inter-metallic that degrades the wirebonds, so the maximum bake time at higher temperatures is much shorter.

The metal routing in the package is 10µm thick tungsten. 
The package surface is gold coated, with a thickness of the metal plating is 2.5µm gold minimum on 1.27µm Nickel minimum.
Aluminum nitride ceramic was chosen, instead of alumina, because of its higher thermal conductivity and to closer match the thermal expansion properties of the silicon device.

The full package stack-up can be seen in Fig.~\ref{fig:phoenix_schematic} or~\ref{fig:peregrine_schematic} for the Phoenix or Peregrine traps, respectively (the top two components are the trap and interposer).

\subsection{Wirebonding}\label{Wirebonding}

\begin{figure}[htbp]
	\includegraphics[width=\textwidth]{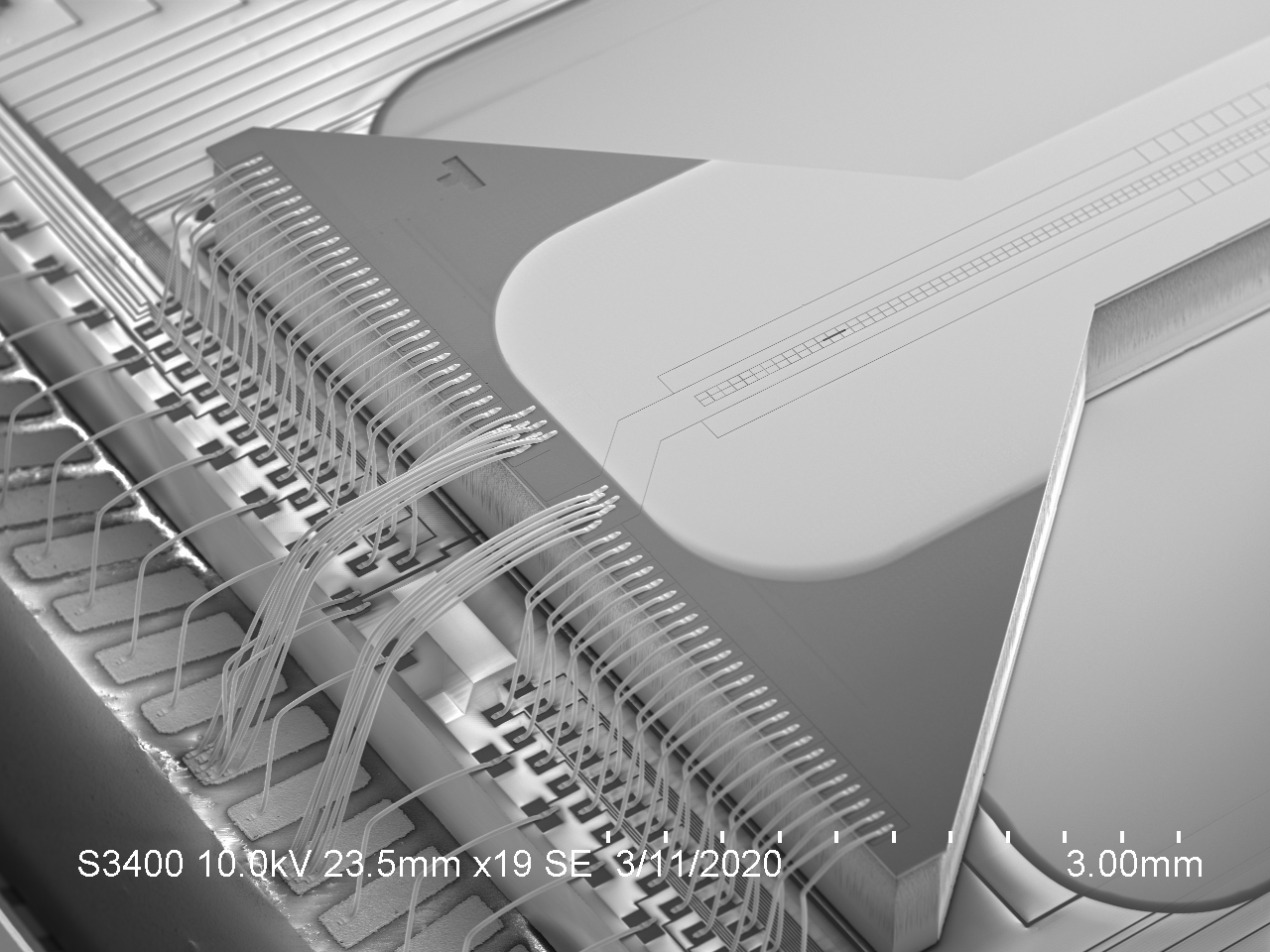}
\caption{\label{fig:wirebonds}Scanning electron micrograph of the wirebonds on a Peregrine trap chip to the package. 
Wirebonds connect all dc control signals from the trap to the package. 
The rf electrode is connected with 7 wirebonds directly to the package to keep the capacitance as low as possible.
There is a gap in the center of the device to allow a beam to pass over the isthmus unobstructed. 
 }
\end{figure}

Wirebonding locations are optimized for beam access.
There are no bondpads down the center isthmus of the trap to allow access for laser beams. 
The trap chip bondpads are located on the long ends of the bowtie and are $95~\um$ wide with a $5~\um$ gap to each neighbor.  
Both the control electrodes and the rf are wire bonded directly to the package.
To allow for higher voltages, the rf is connected to the package with 7 wirebonds, while only 1 is used for each control electrode.
Figure~\ref{fig:wirebonds} shows images of the low profile wirebonds near the rf feed of the trap.

\subsection{Testing}
Before delivery the packaged traps are tested for shorts between every combination of electrodes.  
The capacitance is measured from every control electrode to ground as a test of the integrated trench capacitors.  
Using a charge induced image contrast technique with an SEM the electrode connections are verified, testing the possibility that a disconnected wirebond or internal via leaves the electrode floating.  
The rf electrode resistance is measured to ground with a separate multi-meter to ensure that it exceeds $40\unit{M\Omega}$.

Finally, the device is inspected optically to ensure no surface contaminants are present on the trap which would interfere with trap performance.  

The results of these tests are documented and shipped with the devices, which are labeled with a unique designator on the package surface that also provides internal information regarding the specific wafer and fabrication steps, as well as the photomasks used in fabrication.  
An example of this documentation for a Phoenix trap on a CPGA can be seen in QL139401G\_W17-15.pptx.

If any non-standard features are present (such as wirebond jumpers or additional capacitors), it will be detailed in the part-specific testing documentation.

The devices are maintained in a cleanroom environment throughout their assembly.  
After final electrical testing they are plasma cleaned, with an argon plasma treatment for 5 minutes at 10 Watts, then stored in a package and placed in an antistatic bag for shipment.

\section{Trap Package}

The Phoenix and Peregrine are packaged on a custom designed high-temperature co-fired ceramics (HTCC) package.
The package is mostly backwards compatible with the stock package established during IARPA MQCO program.
A custom package was realized for the Phoenix and Peregrine traps to improve performance and simplify the trap packaging process.

\begin{figure}[ht]
	\centering
	\includegraphics[width=0.45\textwidth]{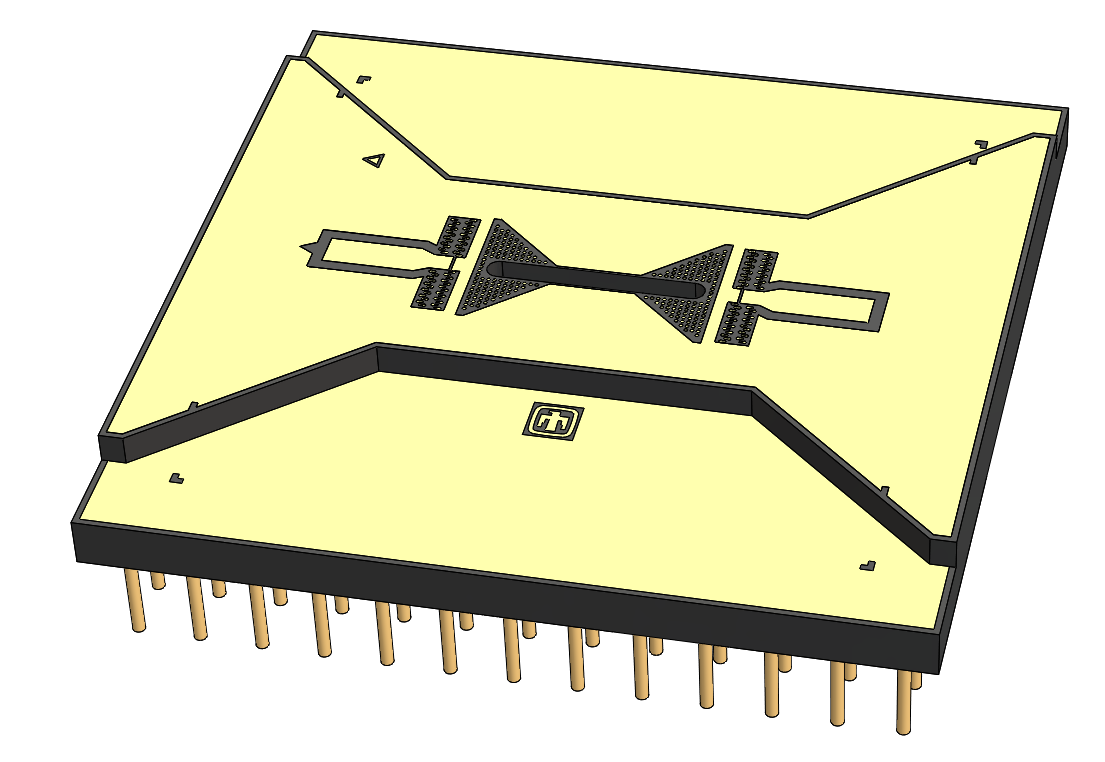}\hfill
	\includegraphics[width=0.45\textwidth]{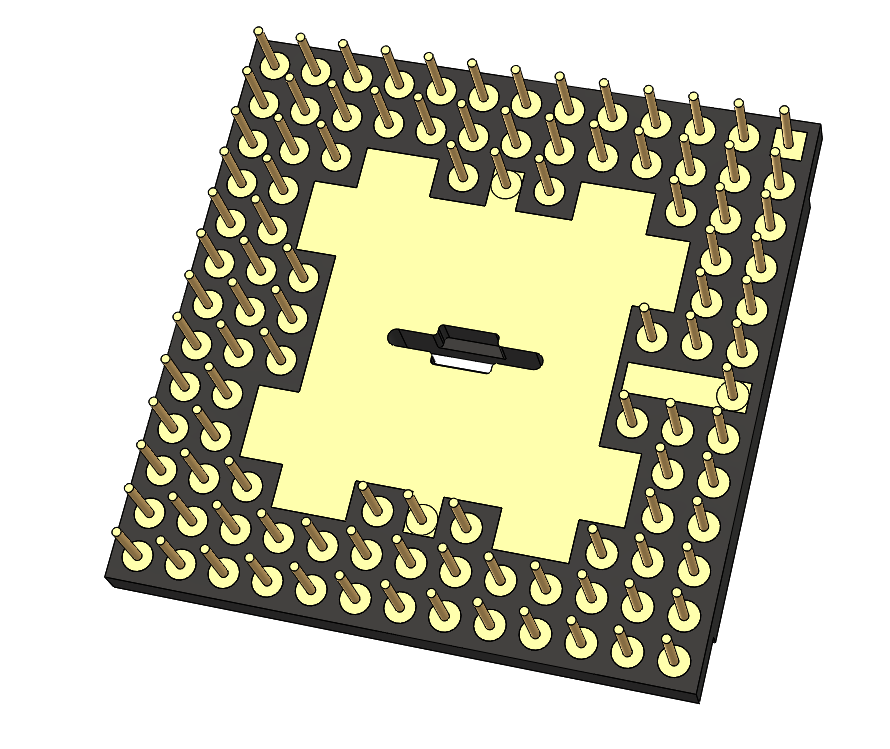}
	\caption{\label{fig:CPGA_package_rendering} Renderings of the CPGA package. 
	The bowtie shape preserves the optical access to the isthmus region of the trap. 
	All metal surfaces of the package are electrically connected to one or more pins (no floating metal). 
	The pins have the same pitch and diameter has the previously used MQCO standard and the pin-out is mostly backwards compatible.
	The small triangle located in the upper left corner of the package bowtie marks the rf feed direction of the device as does the small triangle tab on the rf trace (center left on the package).  }
\end{figure}

\begin{figure}[htbp]
	\centering
	\includegraphics[width=0.49\textwidth]{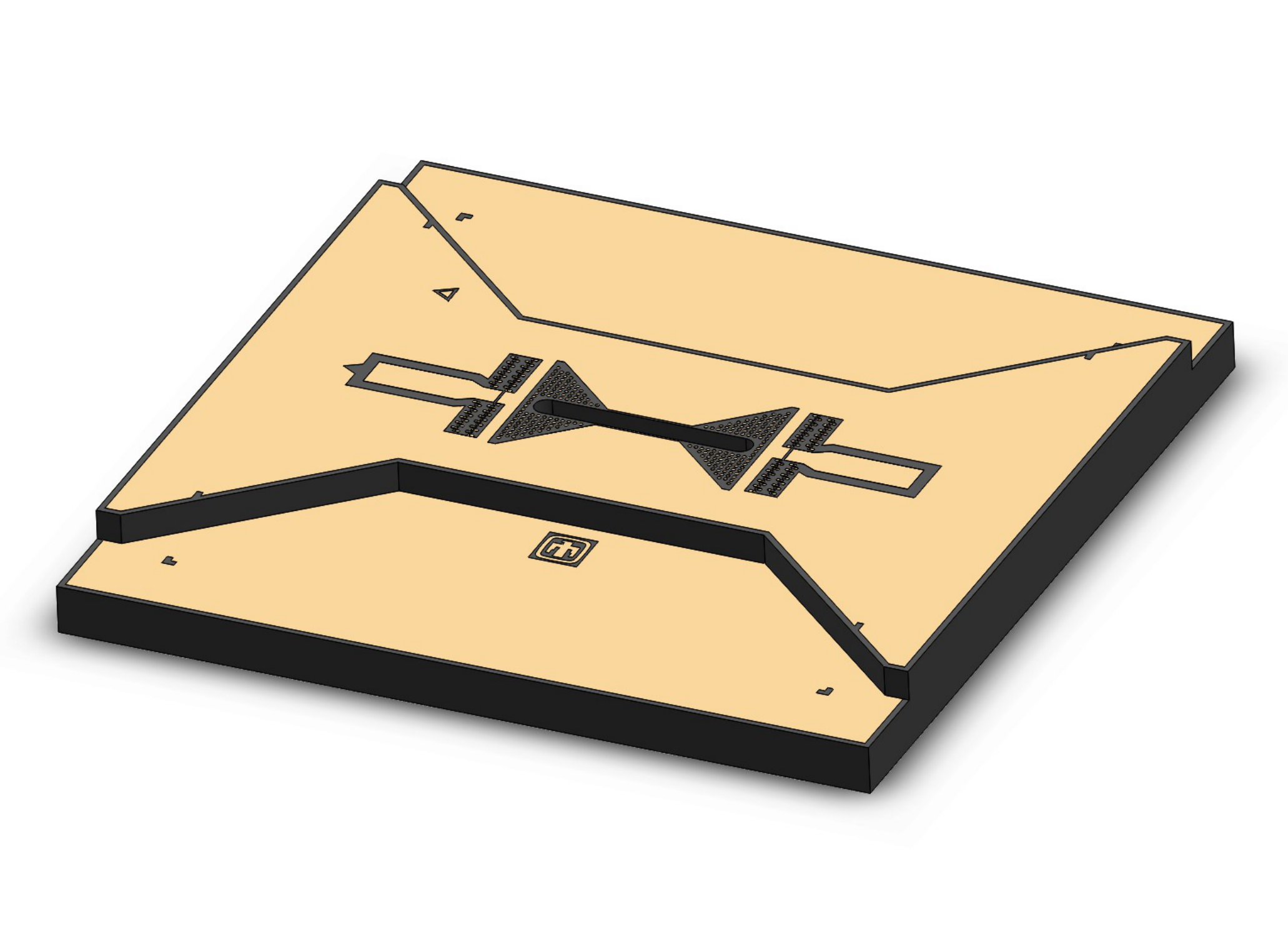}\hfill
	\includegraphics[width=0.45\textwidth]{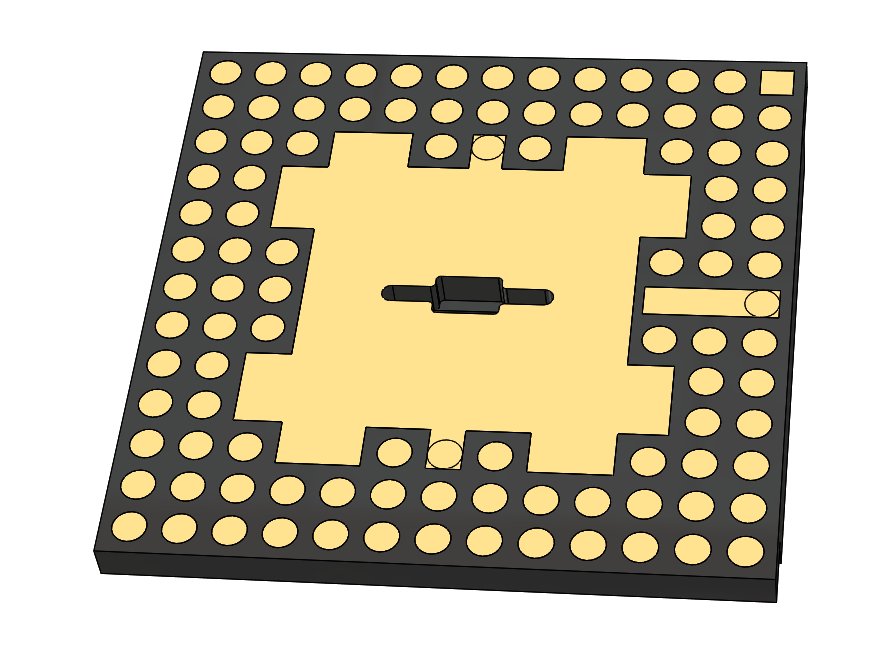}
	\caption{\label{fig:CLGA_package_rendering} Renderings of the CLGA package. 
	The top portion of the package is the same as the CPGA package, but the backside has lands instead of pins. 
	All metal surfaces of the package are electrically connected (no floating metal).}
\end{figure}

\begin{figure}[htbp]
	\centering
	\includegraphics[width=\textwidth]{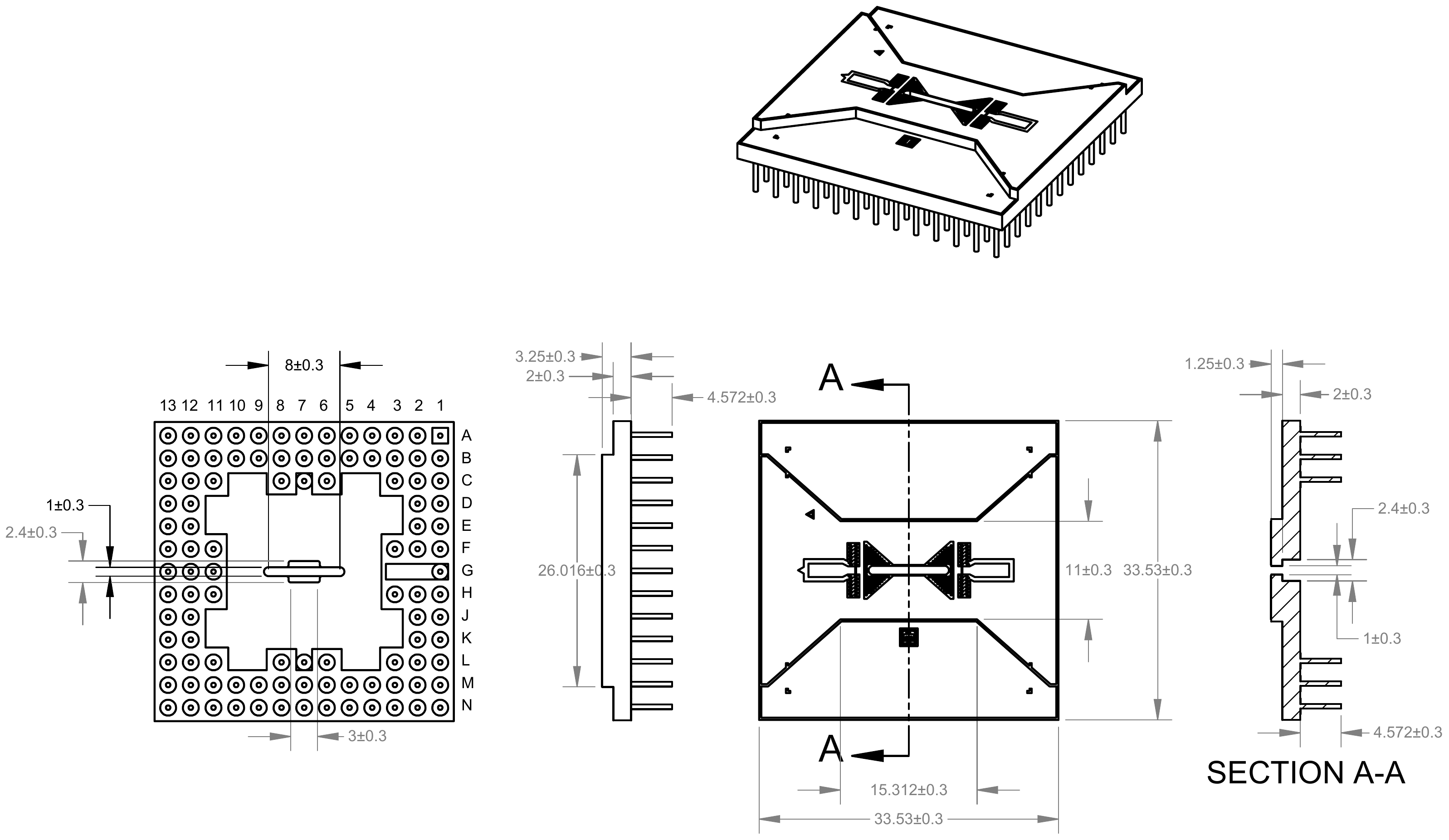}
	\caption{\label{fig:CPGA_package_mechanical} Mechanical dimensions of the CPGA Phoenix package.}
\end{figure}

\begin{figure}[htbp]
	\centering
	\includegraphics[width=\textwidth]{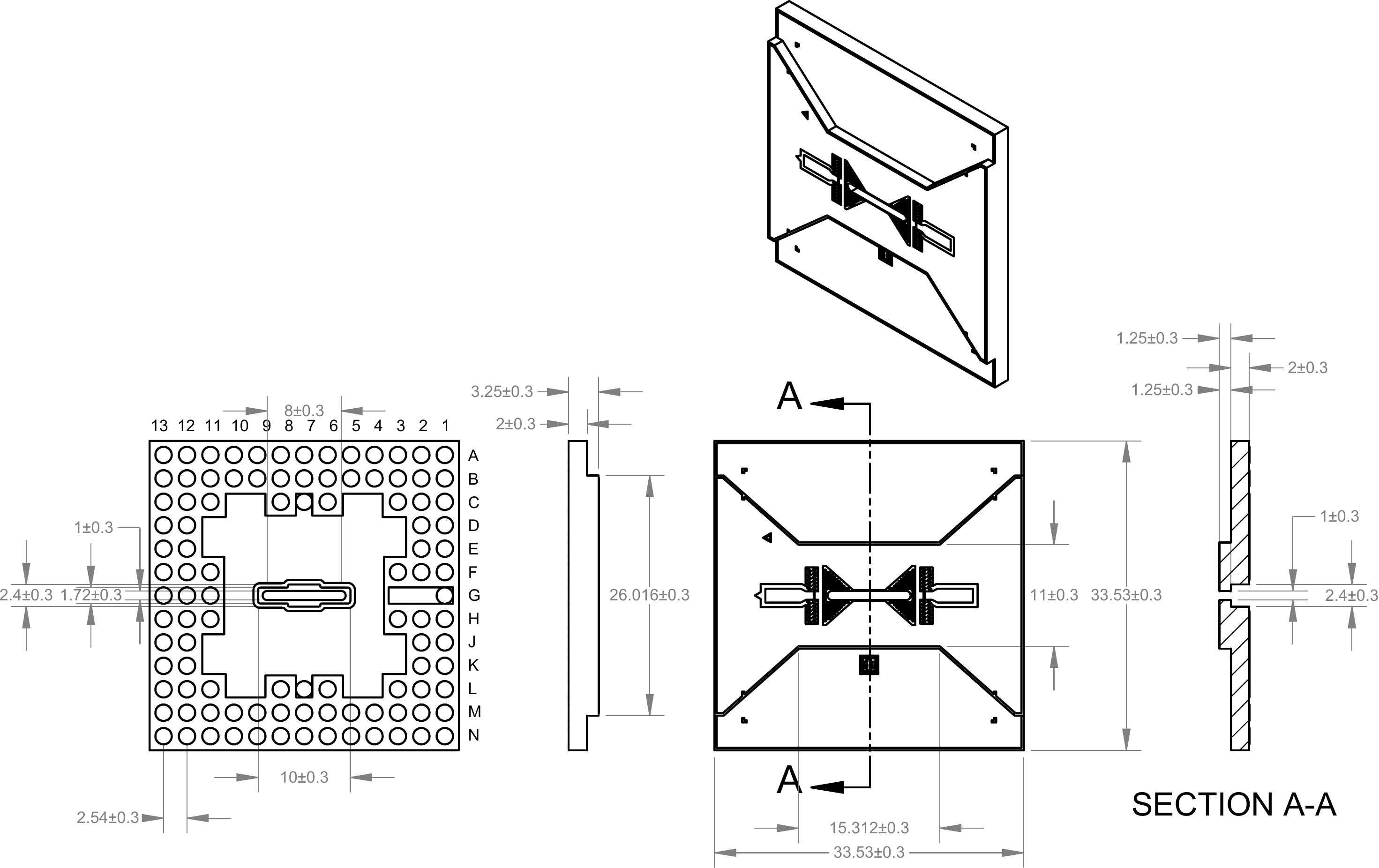}
	\caption{\label{fig:CLGA_package_mechanical} Mechanical dimensions of the CLGA Phoenix package.}
\end{figure}

The package is available with and without pins as Ceramic Pin Grid Array Package (CPGA) and as Ceramic Land Grid Array Package (CLGA), respectively.
A rendering of the CPGA package is shown in Figure~\ref{fig:CPGA_package_rendering}. 
Figure~\ref{fig:CLGA_package_rendering} shows the land grid array package.
The package offers improved performance with respect to thermal conductivity, rf trace resistance, and grounding of the package. 
To realize good thermal conductivity, the package is fabricated from aluminum nitride (AlN). 
Superior rf resistivity is achieved by routing the rf trace on the outside of the package where good conducting materials are available. 
The necessary vias are massively parallel to reduce resistivity. 
All exposed metal on the top and bottom of the package that is not a control signal, is connected to ground.
Close attention was similarly paid to the routing of the rf ground to reduce the resistivity of the rf return signal.
The mechanical dimensions of each package can be found in Figs.~\ref{fig:CPGA_package_mechanical} and ~\ref{fig:CLGA_package_mechanical}, for the CPGA and CLGA, respectively.

The resistance and capacitance of the rf trace were measured to be $\unit[40.8]{m\Omega}$ and $\unit[3.7]{pF}$, respectively. 
The updated package has a lower capacitance of 2.9~pF.

\subsection{Package Netlist}
\begin{figure}
	\centering
	\includegraphics[width=0.6\textwidth]{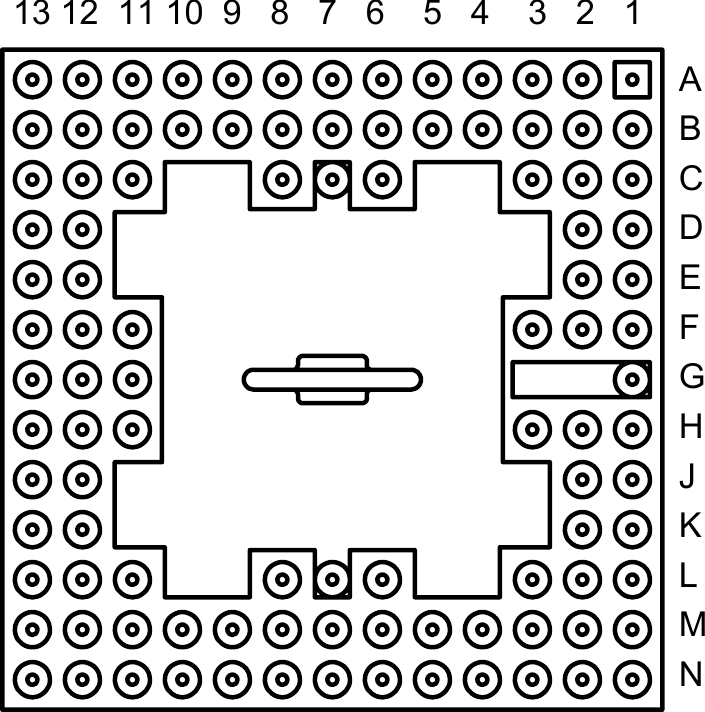}
	\caption{
		\label{fig:cpga_drawing}
		Drawing of the CPGA package as seen from the backside. 
		The main rf connection is at location G1.
		Sites G2 and G3 are also connected to the rf signal; however, these sites do \emph{not} have a pin to enable usage in chambers designed for packages implementing the MQCO packaging standard.
		Ground is at sites C7 and L7.
		Compared to the MQCO packaging standard, this package has 4 additional pins at locations C3, C11, L3, L11. 
		These pins are used for resistive wires on the trap die. 
	}
\end{figure}

The CPGA package (Figure~\ref{fig:cpga_drawing}) has $102$ pins. There is one pin each for rf (G1), and secondary rf (G11) connections,
$2$ ground pins at locations C7 and L7, and $98$ control signal pins.
To achieve lower resistance, the rf signal is routed at the outside of the package from site G1 through site G3.
In the CPGA package the pins at sites G2 and G3 are omitted.
In the CLGA package sites G2 and G3 can optionally be used for additional rf connections.
In addition there are $16$ additional locations that allow for ground connections at locations C4, C5, C9, C10, D3, D11, E3, E11, J3, J11, K3, K11, L4, L5, L9, and L10.    

\begin{figure}
	\centering
	\includegraphics[width=\textwidth]{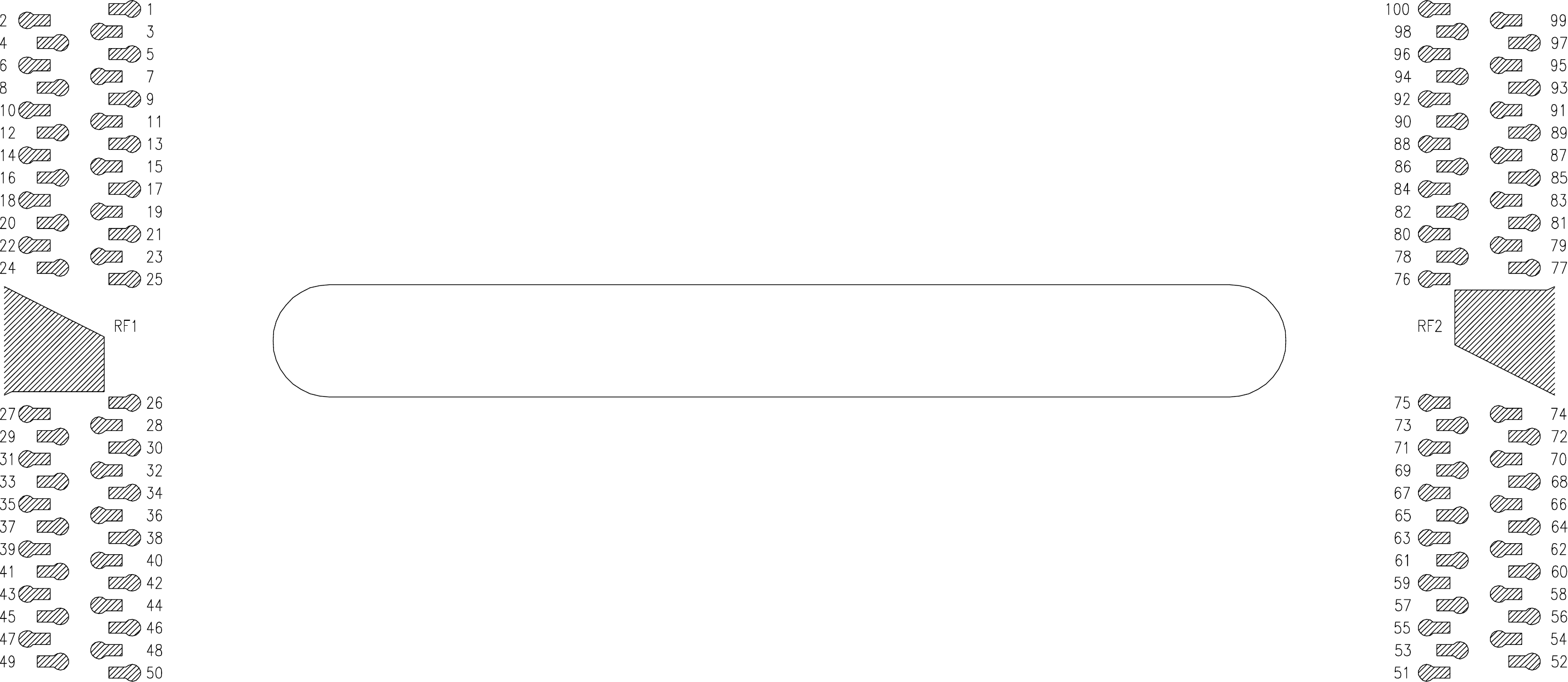}
	\caption{
		\label{fig:package_pads}
		\textbf{Package Pads.} Numbering of the pads of the Phoenix package. 
		The package provides $100$ control signal bondpads, two rf bondpads (RF1 and RF2) as well as a ground plane (not shown here). 	
	}
\end{figure}

The bondpads on the front side of the package (ID-1 \dots ID-100) are labeled in Figure~\ref{fig:package_pads} and the package grid locations are 
labeled in Figures~\ref{fig:CPGA_package_mechanical} and \ref{fig:CLGA_package_mechanical}. 
These are connected according to the following Netlist in Table~\ref{tab:PackageGridID}.

\begin{longtable}{|lc|lc|lc|}
\caption{Netlist for the package grids (from Fig.~\ref{fig:cpga_drawing} to the package bondpads (ID-1 \dots ID-100).}
\label{tab:PackageGridID}\\
	\hline
	ID    & Package Grid & ID    & Package Grid & ID    & Package Grid \\\hline\hline
	\endfirsthead
	\caption[]{Continued.}\\
	\hline
	ID    & Package Grid & ID    & Package Grid & ID    & Package Grid \\\hline\hline
	\endhead
	ID-1  & C3    & ID-35 & H1    & ID-69 & M10 \\
	ID-2  & C7    & ID-36 & N3    & ID-70 & L13 \\
	ID-3  & B2    & ID-37 & M1    & ID-71 & M7 \\
	ID-4  & D1    & ID-38 & N6    & ID-72 & H11 \\
	ID-5  & A5    & ID-39 & J1    & ID-73 & N10 \\
	ID-6  & E2    & ID-40 & M3    & ID-74 & G12 \\
	ID-7  & A2    & ID-41 & L2    & ID-75 & L8 \\
	ID-8  & A1    & ID-42 & M5    & ID-76 & C8 \\
	ID-9  & B5    & ID-43 & J2    & ID-77 & F11 \\
	ID-10 & E1    & ID-44 & N2    & ID-78 & A10 \\
	ID-11 & B3    & ID-45 & N1    & ID-79 & G13 \\
	ID-12 & C2    & ID-46 & N5    & ID-80 & B7 \\
	ID-13 & A6    & ID-47 & K1    & ID-81 & F12 \\
	ID-14 & F1    & ID-48 & M2    & ID-82 & B10 \\
	ID-15 & A3    & ID-49 & L7    & ID-83 & C13 \\
	ID-16 & B1    & ID-50 & L3    & ID-84 & B8 \\
	ID-17 & A7    & ID-51 & L11   & ID-85 & F13 \\
	ID-18 & F2    & ID-52 & K13   & ID-86 & A11 \\
	ID-19 & B4    & ID-53 & M12   & ID-87 & D12 \\
	ID-20 & D2    & ID-54 & N13   & ID-88 & A8 \\
	ID-21 & B6    & ID-55 & N9    & ID-89 & E13 \\
	ID-22 & F3    & ID-56 & J12   & ID-90 & B11 \\
	ID-23 & A4    & ID-57 & N12   & ID-91 & B13 \\
	ID-24 & C1    & ID-58 & L12   & ID-92 & B9 \\
	ID-25 & C6    & ID-59 & M9    & ID-93 & E12 \\
	ID-26 & L6    & ID-60 & J13   & ID-94 & A12 \\
	ID-27 & H3    & ID-61 & M11   & ID-95 & C12 \\
	ID-28 & N4    & ID-62 & M13   & ID-96 & A9 \\
	ID-29 & L1    & ID-63 & N8    & ID-97 & D13 \\
	ID-30 & M6    & ID-64 & H13   & ID-98 & B12 \\
	ID-31 & H2    & ID-65 & N11   & ID-99 & A13 \\
	ID-32 & M4    & ID-66 & K12   & ID-100 & C11 \\
	ID-33 & K2    & ID-67 & M8    & RF    & G1 \\
	ID-34 & N7    & ID-68 & H12   & RF2   & G11 \\
	\hline
\end{longtable}%

\subsection{Trap on Package Netlist}
Phoenix and Peregrine trap share the same electrode names and internal routing; thus, the following Netlist applies to both traps. 
In the following Netlist table, the Package grid locations are labeled in Figure~\ref{fig:cpga_drawing} and the pad locations are in shown in Figure~\ref{fig:package_pads}. 
The electrode names are described in Figure~\ref{fig:phoenix_schematic} and Figure~\ref{fig:peregrine_schematic}. 
The trap die contains $146$ trench capacitors with a capacitance of $\unit[0.8]{nF}$ each.
Electrodes with a larger capacitance to rf are connected to more than $1$ capacitor.
The Capacitor Count column in the following Netlist shows how many capacitors are connected to each signal.

Resistive, and Sense are two aluminum wires located on the penultimate metal layer. 
The Sense wire is connected to two trench capacitors.
The Resistive wire is not connected to trench capacitors and can thus accommodate a larger voltage.
At room temperature the resistance of these wires is approximately 1.75~k$\Omega$.
See Sec.~\ref{sec:tempwire} for more detailed information.

W1 and W2 are two tungsten resistive wires located close to the silicon substrate but electrically separated from it (see Sec.~\ref{sec:tempwire}).
These wires follow the perimeter of the trap die and can be used to either heat the substrate or to estimate its temperature. 
These wires are \emph{not} connected to capacitors.

\begin{longtable}{llll}
\caption{Netlist for the electrodes to the package pins (from Fig.~\ref{fig:cpga_drawing}).}
\label{tab:ElectrodeID}\\
	\hline
	Package Grid & Function & Electrode & Capacitor Count\\ 
	\endfirsthead
	\hline
	Package Grid & Function & Electrode & Capacitor Count\\ 
	\hline
	\endhead
	\hline
	\multicolumn{4}{r@{}}{continued \ldots}\\
	\endfoot
	\hline
	\endlastfoot
	A1    & Signal & Q18   & 1 \\
	A2    & Signal & L0    & 1 \\
	A3    & Signal & Q6    & 1 \\
	A4    & Signal & Q0    & 1 \\
	A5    & Signal & L4    & 1 \\
	A6    & Signal & Q54   & 2 \\
	A7    & Signal & Q8    & 1 \\
	A8    & Signal & Q22   & 1 \\
	A9    & Signal & Q40   & 1 \\
	A10   & Signal & Q38   & 1 \\
	A11   & Signal & Q26   & 1 \\
	A12   & Signal & Q32   & 1 \\
	A13   & Signal & W1    &  \\
	B1    & Signal & Q50   & 2 \\
	B2    & Signal & L8    & 1 \\
	B3    & Signal & Q12   & 1 \\
	B4    & Signal & Q4    & 1 \\
	B5    & Signal & Q16   & 1 \\
	B6    & Signal & Q46   & 2 \\
	B7    & Signal & Q62   & 2 \\
	B8    & Signal & Q30   & 1 \\
	B9    & Signal & Q24   & 1 \\
	B10   & Signal & Q34   & 1 \\
	B11   & Signal & S6    & 2 \\
	B12   & Signal & W1    &  \\
	B13   & Signal & Q20   & 1 \\
	C1    & Signal & S8    & 2 \\
	C2    & Signal & Q10   & 1 \\
	C3    & Signal & SENSE & 2 \\
	C6    & Signal & Q44   & 2 \\
	C7    & GND   & GND   &  \\
	C8    & Signal & Q64   & 2 \\
	C11   & Signal & SENSE & 2 \\
	C12   & Signal & Q36   & 1 \\
	C13   & Signal & Q60   & 2 \\
	D1    & Signal & L6    & 1 \\
	D2    & Signal & Q2    & 1 \\
	D12   & Signal & Q56   & 2 \\
	D13   & Signal & Q42   & 1 \\
	E1    & Signal & Q14   & 1 \\
	E2    & Signal & L2    & 1 \\
	E12   & Signal & Q28   & 1 \\
	E13   & Signal & S4    & 2 \\
	F1    & Signal & Q52   & 2 \\
	F2    & Signal & Q48   & 2 \\
	F3    & Signal & S10   & 2 \\
	F11   & Signal & O0    & 10 \\
	F12   & Signal & S0    & 2 \\
	F13   & Signal & Q58   & 2 \\
	G1    & RF    & rf    &  \\
	G11   & RF    & rf\_sense &  \\
	G12   & Signal & O1    & 10 \\
	G13   & Signal & S2    & 2 \\
	H1    & Signal & Q51   & 2 \\
	H2    & Signal & Q3    & 1 \\
	H3    & Signal & S9    & 2 \\
	H11   & Signal & S3    & 2 \\
	H12   & Signal & Q61   & 2 \\
	H13   & Signal & Q57   & 2 \\
	J1    & Signal & Q11   & 1 \\
	J2    & Signal & Q19   & 1 \\
	J12   & Signal & Q37   & 1 \\
	J13   & Signal & Q21   & 1 \\
	K1    & Signal & L7    & 1 \\
	K2    & Signal & Q49   & 2 \\
	K12   & Signal & Q59   & 2 \\
	K13   & Signal & W2    &  \\
	L1    & Signal & S11   & 2 \\
	L2    & Signal & Q15   & 1 \\
	L3    & Signal & Resistive &  \\
	L6    & Signal & Q45   & 2 \\
	L7    & GND   & GND   &  \\
	L8    & Signal & Q65   & 2 \\
	L11   & Signal & Resistive &  \\
	L12   & Signal & Q29   & 1 \\
	L13   & Signal & S1    & 2 \\
	M1    & Signal & Q53   & 2 \\
	M2    & Signal & L9    & 1 \\
	M3    & Signal & Q13   & 1 \\
	M4    & Signal & Q5    & 1 \\
	M5    & Signal & Q17   & 1 \\
	M6    & Signal & Q47   & 2 \\
	M7    & Signal & Q63   & 2 \\
	M8    & Signal & Q31   & 1 \\
	M9    & Signal & Q25   & 1 \\
	M10   & Signal & Q35   & 1 \\
	M11   & Signal & S7    & 2 \\
	M12   & Signal & W2    &  \\
	M13   & Signal & S5    & 2 \\
	N1    & Signal & L3    & 1 \\
	N2    & Signal & L1    & 1 \\
	N3    & Signal & Q7    & 1 \\
	N4    & Signal & Q1    & 1 \\
	N5    & Signal & L5    & 1 \\
	N6    & Signal & Q55   & 2 \\
	N7    & Signal & Q9    & 1 \\
	N8    & Signal & Q23   & 1 \\
	N9    & Signal & Q41   & 1 \\
	N10   & Signal & Q39   & 1 \\
	N11   & Signal & Q27   & 1 \\
	N12   & Signal & Q33   & 1 \\
	N13   & Signal & Q43   & 1 \\
\end{longtable}%

\section{Trap Packaging and Parametric Testing}	

While bare die are available, packaging the traps makes it possible to handle and ship the trap easily and keeps mounting the trap in a vacuum chamber straightforward. 
Additionally, once the traps are mounted on the package, they can be parametrically tested for shorts and correct capacitance. The traps are also tested for connectivity between the trap electrodes and the pins or lands.
The Phoenix or Peregrine traps are solder die attached to the Phoenix package using gold-tin (AuSn) solder.
Introduction of this solder die attach process means that trap assemblies do \emph{not} contain epoxies and are completely organics-free.

\section{Trap Installation}
The following describes the installation procedure for the user after receiving the trap.

Microfabricated surface traps are sensitive to dust accumulation on the surface. At Sandia, the traps are only handled in cleanrooms. The traps are packaged in a cleanroom and should {\em only be opened in a cleanroom.}

The Phoenix trap with integrated trench capacitors can be damaged by static electricity. Voltages above $\pm 30\unit{V}$ will damage the trench capacitors attached to all control electrodes.

\subsection{Vacuum Chamber}
The user is responsible for building a UHV vacuum chamber to accommodate the packaged device.  A chamber with $94$ control voltages wired to be compatible with the package is necessary for the successful operation of the trap. 

The success of these experiments depends on achieving an excellent vacuum.  All surfaces exposed to the interior of the chamber need to be handled in an absolutely grease-free way.  Pressures below $10^{-10}~\unit{mbar}$ should be acceptable for many ion trapping experiments, although pressures significantly lower will result in superior trapping times (since the trap depth is only a few times room temperature). 

In addition, the traps are sensitive to dust particles accumulated on the surface, as they can cause electrical shorts, accumulate static charges, and cause laser scatter.  Therefore, we strongly recommend installing the traps in a cleanroom environment to prevent contamination. 

The orientation of the trap in the vacuum chamber has an influence on the likelihood of getting dust particles located on the trap surface. If the trap surface is vertical or pointing down, dust accumulation on the surface is less likely.

Vacuum compatible zero insertion force (ZIF) sockets for the 100 pin CPGA package are available from Tactic Electronics(PN: 100-4680-001A).

\subsection{Ground Plane Above Trap}
The user is responsible for mitigating charge on nearby dielectric surfaces.  
Generally, the closest and most influential surface is the imaging viewport.  
While the package itself is not much farther away from the ion, the geometry limits the impact of charge that may build up on that surface (since there are many screening grounded regions between exposed dielectric of the package and the ion).  
The viewport, however, has direct line-of-sight to the ion and unless mitigated is mostly unscreened.  Possible approaches include:

\begin{enumerate}
	\item  Building a custom metal screen which accommodates the necessary laser and imaging access and mounting this structure on the package. Screens with openings up to an NA of 0.6 have been operated successfully.
	
	\item  Coating the re-entrant viewport with a conductive transparent material (such as ITO)
	
	\item  Using a mesh with a small geometric fill factor.
\end{enumerate}

In addition, the screen should be placed far enough away as to not have a significant impact on the electric field for an otherwise unscreened trap, which is how they are simulated.  
A rule of thumb is that if the distance of the screen to the trap surface is more than $20$ times the distance of the ion above the trap surface $(\approx 70\um)$, then the presence of the ground will have minimal impact on the trap behavior.  
Above this separation, the specific distance does not significantly affect operating conditions, such as the trap strength for a given rf voltage or the control waveforms.  
A $5~\unit{mm}$ distance between a grounded mesh and the trapping surface has worked well on experiments at SNL. 
For screens closer than 5~mm, the voltage solutions may have to be adjusted for the presence of the ground screen.

\subsection{Trap Insertion}

These traps have to be installed with the gold tab on the package pointed in the direction of the socket's rf pins.  
The packages are quite robust and can be pushed in with a great deal of pressure, however the user should be careful not to touch the wirebonds on the trap and only contact the package surface. 
Also make sure that none of the pins are being bent while inserting the package into the socket.
If the trap needs to be retracted and used again, care should be taken to not bend the package pins.

The traps have to be installed in a dust-free and clean environment, ideally a cleanroom, to limit the possibility of dust falling on the chip surface.
If that is not feasible, they should be installed as quickly as possible to minimize exposure.

\subsection{Pre-bake Testing}
\label{sec:prebake}
Before baking the user should verify that the rf electrode is not shorted to ground and measure its capacitance.  
The capacitance of the packaged device should be $11 \unit{pF}$, $7\pm 1 \unit{pF}$ for the trap die ($5.5\pm 1 \unit{pF}$ for the Peregrine) and $4\pm 1 \unit{pF}$ for the CPGA carrier.  
This does not include the capacitance of the feedthrough and socket (which depends on the wiring but is $\approx 6 \unit{pF}$ in our experiment) and should be measured by the user before device insertion. 
If higher than normal voltages are to be applied, this should be tested at high vacuum before baking to make sure the device can handle them. 
Tests can be done with a dc voltage which should be applied through a large resistor to limit the current. 
The leaking current should be measured with a picoamp meter. 
In this setup, a rising current is indicative of a voltage close to breakdown.

The user should also verify that all dc control electrodes are not shorted to ground and measure their capacitance; if the capacitance is a multiple of 0.8~nF when it shouldn't be, the electrode is likely shorted to a neighbor (either on the chip or in the wiring of the vacuum chamber).

\subsection{Baking}
The devices themselves can be baked up to $200^{\circ}C$, limited by the die attach and the development of purple plague~\cite{chen_physics_1967}. 
It may be the case that other components (such as Kapton wires) in the vacuum chamber limit the bake to a lower temperature.

As described in section~\ref{Wirebonding}, the device is packaged using wirebonding with gold wires connecting to aluminum pads. 
If this junction is heated, brittle aluminum-gold intermetallics are formed. The aluminum-gold intermetallic will lead to a slightly higher contact resistance and, most importantly, to a reduced pull strength of the wirebonds. 
This process depends strongly on the baking temperature (see Figure~2 in ~\cite{chen_physics_1967}), thus limiting the baking time at high temperatures is most effective in preventing the formation of purple plague. 
Our tests show that baking of up to $7~\unit{days}$ at $200\unit{^\circ C}$ is acceptable.

\section{Trap Operation}
\subsection{Rf Voltage Application}
(All capacitance values are estimates and subject to change between devices.)
The Phoenix devices have a total capacitance between rf high and ground of $\unit[7]{pF}$, including the package (which adds $\unit[4]{pF}$).  
Depending on the vacuum chamber, the feedthrough, internal rf wiring, and socket can add an additional $\approx\unit[6]{pF}$, bringing the total capacitive load to $\approx\unit[15]{pF}$.  

In an ideal quadrupole trap, the trap secular frequency is given by:
\[{\omega }_{\text{sec}}=\frac{qV}{\sqrt{2}{\Omega }_{\text{rf}}\, mR^2}, \] 
where $q$ is the charge of the trapped particle, $V$ is the voltage amplitude, $\Omega_{\text{rf}}$ is the applied rf angular frequency, $m$ is the mass of the particle, and $R$ is electrode distance.  This can be simply modified to calculate the secular frequency in the Phoenix trap by substituting $R$ for the characteristic distance of the device. The characteristic distance corresponds to the distance at which hyperbolic electrodes operating at the same voltage and frequency would generate the same secular frequency as in the surface trap. The characteristic distance for the Phoenix trap in the slotted region is $$R=140\um.$$ 
By substituting $R$ with this number the user can calculate the secular frequency they are generating for a particular voltage, drive frequency, and ion species.

The trap depth can be expressed as:
$$d=\alpha \frac{1}{2}m{\omega }^2_{\text{sec}} R^2,\ \ \alpha =0.028, $$ 
where the factor $\alpha $ is required to account for the specifics of the surface trap geometry.  In terms of practical trap depth (defined as the ability to stay trapped following a collision with a particle of energy $E$, this assumes that the stability parameter $h$ does not exceed $0.5$.  

\subsubsection{Rf Voltage Limit}
\label{sec:Vlimit}

To improve the probability of a good fabrication yield, the separation between the rf and ground was decreased in these traps as compared to previous traps fabricated by Sandia. 
The vertical separation between rf and ground in the Phoenix and the Peregrine traps is $2~\um$ (as compared to $10~\um$ on an HOA-2 trap). 
This decrease in separation has decreased maximum safe voltage.
This value is dependent on the thickness of the gold applied to the trap and ultimately, the maximum is different for each device and would need to be verified by the user. 

For thin gold (Ti-100~nm/Pt-100~nm/Au-250~nm), we have measured a typical dc voltage breakdown of about 350~V. 
This corresponds to an immediate rf breakdown voltage of approximately 250~V (voltage amplitude). 
Meaning, application of voltages that high will immediately develop a short.
Safe long-term operation of the traps can occur at voltages of about 150~V, corresponding to maximum frequencies of about 2.0~MHz. 
The dc breakdown voltage shows an increase in maximum voltage when the trap is operated at cryogenic temperatures, however, the rf tests are inconclusive.
For now, we do not encourage operating the trap at higher voltages unless you are willing to test the device and risk developing a short. 

Traps with thick gold (Ti-100~nm/Pt-100~nm/Au-500~nm) have demonstrated an even lower dc breakdown voltage, around 220~V. 
This suggests an immediate rf breakdown point of 160~V and a long term rf safe value of about 100~V, corresponding to maximum trap frequencies of about 2.3 and 1.6~MHz respectively. 
While thick gold traps are typically argon-ion plasma cleaned, we do not know how much gold is removed from the lower metal layers to give a good estimate of the increase in breakdown voltage that cleaning may provide. 

The method for testing the dc voltage limit is described in Section~\ref{sec:prebake}. 
This gives a good estimate for the rf voltage limits. 
In the case where the signs of dc breakdown were not obvious and the trap showed evidence of arcing as the first indication of failure, we have found this will lower the dc breakdown point, but does not seem to change the rf breakdown, at least not significantly on a single arcing event. 
Based on a small sample, an empirical relationship between the measured dc breakdown and the safe rf running configuration is $V_{rf~ safe} \approx V_{dc~ break}*0.43$. 
This holds true as long as the testing conditions were similar to the running conditions (meaning the vacuum pressure and the device temperature). 

\subsubsection{Remove a Short}

\begin{figure}
	\centering
	\includegraphics[width=0.45\textwidth]{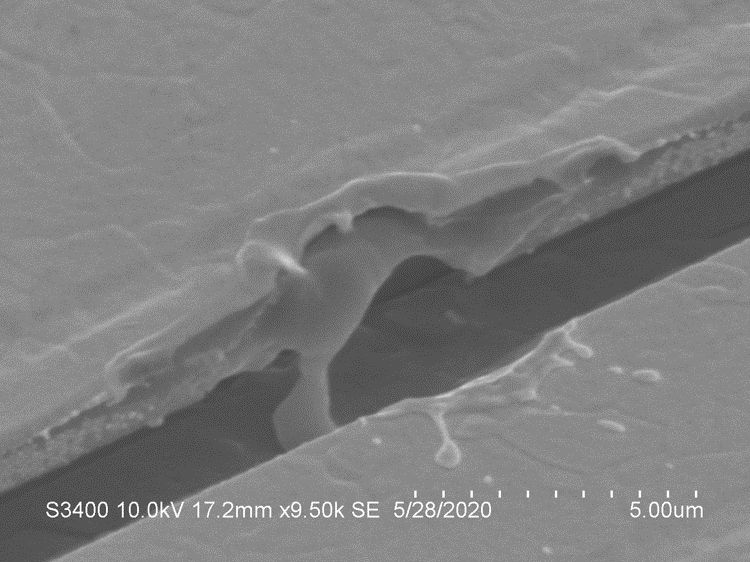}
	\caption{
		\label{fig:short}
		\textbf{SEM image of a Phoenix trap short from rf to ground.}
		A small (about $0.5~\um$) filament developed at one of the arcing locations. 
		This trap had 7 arcing events that left visible marring on the metal, only this location shorted the rf. 
			}
\end{figure}

If a short has developed on the trap due to this breakdown mechanism, the rf should measure a very low resistance, typically about 1.5~$\Omega$ using a standard multimeter. 
As the rf begins to arc to the lower ground metal at higher voltages, the metal becomes molten and is thrown off by the arc, see Figure~\ref{fig:short}.
In traps we have tested to study the breakdown mechanism, there are typically many arc events that leave visible evidence, but only 1 short. 
Since the arcing is not correlated to the ion trapping location (as far as we can tell), the probability of disturbing the metal at the trapping site is no higher than anywhere else. 

If the trap does short, the short can be successfully melted away. 
We have had three instances of shorting on devices that are being used to trap ions. 
In each case, we removed the rf resonator and attached dc leads directly to the rf and ground. 
We slowly increased the current going through the rf until we reached approximately 0.5~A. 
At this point, the filament melted and the short opened. 
Afterwards, the device was able to trap again and returned to normal operation. 
As long as the rf voltage was kept below the safe running limit from Section~\ref{sec:Vlimit}, we have yet to develop another short.

\subsection{Control voltages}

Control voltages up to $\pm20~$V are safe. 
The Phoenix/Peregrine trap has a $\pm30~$V breakdown limit for each control channel. 
Exceeding this limitation is very likely to result in a short to ground in the trench capacitor for that channel.  
A current-voltage characteristic corresponding to a diode is indicative of a short between metal and silicon. 
A strictly ohmic short is usually caused by direct metal to metal shorts either on chip or in the wiring.

Control electrodes require capacitors to shunt rf signals to ground, which couple via the parasitic capacitance between the rf electrode and the control electrodes.
The trench capacitors are incorporated directly into the trap chip for each control channel and are designed to be operated at up to $\pm 20~\unit{V}$ and have a breakdown voltage of $> \pm 30~\unit{V}$.  
Each capacitor has a capacitance $C = 0.8~\unit{nF}$, with a variance $<1\%$. 
The stray inductance of the trench capacitors themselves, which can undermine the ability to shunt the rf off the control electrodes, is $0.05~\unit{nH}$, much smaller than the $\approx 1~\unit{nH}$ inductance that the wirebonds between the trap and the interposer add.  
The series resistance of the trench capacitors ($4~\unit{\Omega}$) is comparable to the lead resistance of the routing between bondpad and electrode on the trap chip.

The Phoenix/Peregrine traps have increased control in the quantum region, now offering segmented pairs on both the inner and the outer electrodes in the quantum region. 
The schematic in Fig.~\ref{fig:phoenix_schematic} shows the electrode layout for the Phoenix device (see Fig.~\ref{fig:peregrine_schematic} for the corresponding Peregrine layout).
The rf trace is coming from the left, defining the orientation of the device.
The orientation can also be determined from the triangle on the package (see Fig.~\ref{fig:CPGA_package_rendering} and~\ref{fig:CLGA_package_rendering}) which is located on the left of the package, with the rf feed, and points towards the feed direction.

Table~\ref{tab:PackageGridID} and~\ref{tab:ElectrodeID} list the mapping between the electrode labels for the Phoenix/Peregrine traps, the corresponding ID's, and the package grids. 
The mapping for the Phoenix and Peregrine is identical.
These tables should be combined with the wiring netlist for the particular vacuum system in use for trapping to generate the electrode-to-control channel netlist.

The axial secular frequency generated by a ``unit set'' of control voltages is:
\[{\omega }_{axial}=\sqrt{\frac{D\ U\ q}{m\ }} ,\] 
where $D$ depends on the particular unit voltage set (the geometry, electrodes used, etc.), $U$ is the scale of the voltage applied, $q$ is the charge of the ion, and $m$ is the mass of the ion. 
This formula can be used to determine the effect of scaling the waveforms provided in Section~\ref{sec:ContrlSolGen}. 
The secular frequencies in the radial directions will be impacted by the addition of the control voltages.

\section{Electrostatic Simulation}
\label{sec:waveforms}

\subsection{Trap Simulations}
Boundary element simulations are employed to calculate the charge on all trap electrodes. From the boundary element solutions, the electrostatic potential $U_i(\mathbf{x})$ at location $\mathbf{x}$ generated by an electrical potential of $1\unit{V}$ on electrode $i$ while all other electrodes are at ground potential is calculated. Using the superposition principle, the potential generated by any set of voltages on the trap electrodes can be calculated from the set of $U_i$ potentials. In addition, the ponderomotive potential generated by the rf electrodes can be calculated using 
$$U_\text{pond} = \frac{e V_\text{rf}^2}{4 m \omega_\text{rf} } \nabla U_\text{rf} \cdot \nabla U_\text{rf},$$
where $e$ is the electron charge, $V_\text{rf}$ the voltage amplitude of the radio-frequency drive, $m$ is the ion mass and $U_\text{rf}$ is the electrostatic potential generated by the radiofrequency electrode.

The meshing used for this simulation is documented in \texttt{.msh} and \texttt{.vtk} files. 
The \emph{Visualization Toolkit} files (\texttt{.vtk}) can be displayed using the freely available Paraview software.
The \texttt{.msh} mesh files can be viewed using the also freely available Gmsh software.

\subsubsection{Voltage Arrays}
The values of $U_i(\mathbf{x})$ are calculated near the rf nodal lines for all the electrodes and saved in a file. We call these data Voltage Arrays (\texttt{.va}). 

\paragraph{File format}
The Voltage Arrays are saved in binary format and usually have a \texttt{.va} file extension. A file header giving information about the number of electrodes, boundaries, axes is followed by an array of potential values. The full file format is specified in Table~\ref{table_voltage_array_fileformat}.

\begin{table}
	\begin{tabular}{|l|l|p{6cm}|}
		\hline
		\textbf{Type} & \textbf{Number} & \textbf{Content}\\\hline
		64 bit Integer & 1 & \\\hline
		64 bit Integer & 1 & $N_\text{elec}$: Number of electrodes\\\hline
		64 bit Integer & 3 & $n_x, n_y, n_z$: Number of samples along the $x$, $y$ and $z$ axis\\\hline
		64 bit Integer & 1 & Number of voltage sets (1)\\\hline
		64 bit double & 3 & Re-mapped x-axis\\\hline
		64 bit double & 3 & Re-mapped y-axis\\\hline
		64 bit double & 3 & Stride along x-, y-, and z-axis\\\hline
		64 bit double & 3 & Origin\\\hline
		64 bit Integer & 2 &\\\hline
		64 bit Integer & $N_\text{elec}$ & Electrode mapping\\\hline
		64 bit double & $N_\text{elec} \cdot n_x \cdot n_y \cdot n_z$ & Potential data\\\hline
	\end{tabular}
	\caption{\label{table_voltage_array_fileformat}File format of the binary voltage array files. (for example \texttt{Phoenix.va}) }
\end{table}

\paragraph{Electrode assignment}
The voltage array files contain the potential for the electrodes in the order given in Table~\ref{table_electrodes}.

\begin{table}\centering
	\begin{tabular}{|l|l|}\hline
		\textbf{Index} & \textbf{Electrode}\\\hline
		1 & Ground\\\hline
		2 & rf\\\hline
		$3\dots 14$ & S0 $\dots$ S11\\\hline
		$15 \dots 24$ & L0 $\dots$ L9\\\hline
		$25 \dots 90$ & Q0 $\dots$ Q65\\\hline
		$91 \dots 92$ & O0 $\dots$ O1\\\hline
	\end{tabular}
	\caption{\label{table_electrodes} One-based index (as used in Mathematica) of the electrodes of the Phoenix trap in voltage array files.}
\end{table}

\section{Specifications and Absolute Maximum Ratings}
\begin{longtable}{|p{0.63\textwidth}|p{0.3\textwidth}|}
	\hline \xrowht{11pt}
	\textbf{Geometry} & \\ \specialrule{.15em}{.2em}{.2em}
	Bowtie length & $\unit[10]{mm}$ \\\hline
	Bowtie width & $\unit[6.4]{mm}$ \\\hline
	Bowtie height & $\unit[790]{\mu m}$ \\\hline
	Isthmus width & $\unit[1200]{\mu m}$\\\hline
	Isthmus length & $\unit[3.48]{mm}$ \\\hline
	Slot width (Phoenix Trap)& $\unit[60]{\mu m}$ \\\hline
	Slot length at full width (Phoenix Trap)& $\unit[2.88]{mm}$ \\\hline
	Slot length accessible at $\unit[45]{^\circ}$ & $\unit[2.2]{mm}$ \\\hline
	Backside slot width & $\unit[400]{\mu m}$ \\\hline
	Backside slot length & $\unit[3.62]{mm}$\\\hline
	\hline
	\xrowht{11pt}
	\textbf{Rf Electrode} & \\ \specialrule{.15em}{.2em}{.2em}
	Maximal voltage on rf electrodes & $\pm \unit[180]{V_{pk}}$ \\\hline
	Capacitance of Phoenix trap chip & $\approx \unit[7.1]{pF}$ \\\hline
	Capacitance of Peregrine trap chip & $\approx \unit[5.4]{pF}$ \\\hline
	Capacitance of Phoenix assembly & $\approx \unit[10]{pF}$ \\\hline
	Capacitance of Peregrine assembly & $\approx \unit[8.4]{pF}$ \\\hline
	Est. rf dissipation of Phoenix trap for $\unit[100]{V}$, $\unit[100]{MHz}$ & $\unit[35]{mW}$ \\\hline
	Est. rf dissipation of Peregrine trap for $\unit[100]{V}$, $\unit[100]{MHz}$ & $\unit[20]{mW}$ \\\hline
	Rf sense divider ratio of Phoenix (unloaded)& $\nicefrac{\unit[.25]{pF}}{\unit[60]{pF}} \approx 240$\\\hline
	Rf sense divider ratio of Peregrine& $\nicefrac{\unit[.17]{pF}}{\unit[68]{pF}} \approx 350$\\\hline\hline
	\xrowht{11pt}
	\textbf{DC Control electrodes} & \\\specialrule{.15em}{.2em}{.2em}
	Maximal voltage on dc electrodes & $\pm 30\,\unit{V}$ \\\hline
	Maximal leakage current through trench capacitor at $\pm 10\,\unit{V}$ & 10\,\unit{nA} (non-ohmic) \\\hline 
	Number of control voltages needed & $94$ \\\hline\hline
	\xrowht{11pt}
	\textbf{Quantum region} & \\\specialrule{.15em}{.2em}{.2em}
	Length & $22\times \unit[70]{\mu m} = \unit[1.54]{mm}$ \\\hline
	Inner control electrode pairs & $22$ \\\hline
	Inner control electrode pitch & $\unit[70]{\mu m}$ \\\hline
	Outer control electrode pairs & $11$ \\\hline
	Outer control electrode pitch & $\unit[140]{\mu m}$ \\\hline
	Ion height - Phoenix & $\unit[68]{\mu m}$ \\\hline
	Ion height - Peregrine & $\unit[72]{\mu m}$ \\\hline \xrowht{10pt}
	 & single $x=\unit[0]{\mu m}$\\\xrowht{9pt}
	Marker locations &	double $x=\pm \unit[420]{\mu m}$\\\xrowht{10pt}
	&	loading~$x=\unit[-3045]{\mu m}$ \\\hline
	\hline
	\xrowht{11pt}
	\textbf{Transition between slot and above-surface} &\\\specialrule{.15em}{.2em}{.2em}
	Control electrode pairs (co-wired with shuttling regions) & $6$ \\\hline
	Control electrode pitch & $\unit[70]{\mu m}$ \\\hline
	Maximum curvature trace variation & $\unit[1]{\%}$ \\\hline
	Maximum axial trap frequencies (from rf) relative to radial trap frequencies & $\unit[8]{\%}$  \\\hline
	Ion height Phoenix in slotted part (above top metal) & $\unit[68]{\mu m}$ \\\hline
	Ion height Phoenix above surface part (above top metal) & $\unit[72]{\mu m}$ \\\hline
	Ion height Peregrine & $\unit[74]{\mu m}$ \\\hline
	\hline
	\xrowht{11pt}
	\textbf{Trench capacitors} (all dc control electrodes)& \\\specialrule{.15em}{.2em}{.2em}
	Capacitance & $0.8\unit{nF}$ \\\hline
	Series inductance & $\approx 50\unit{pH}$ \\\hline
	Series resistance ($20\unit{^\circ C})$ & $4 \unit{\Omega}$ \\\hline
	Maximum voltage & $\pm 30\unit{V}$ \\\hline
	Damage Threshold voltage & $> \pm 32 \unit{V}$ \\ \hline\hline
	\xrowht{11pt}
	\textbf{Loading region}&\\\specialrule{.15em}{.2em}{.2em}
	Loading slot length & $\unit[140]{\mu m}$ \\\hline
	Loading slot width (top $\unit[45]{\mu m}$) & $\unit[13]{\mu m}$ \\\hline
	Loading slot backside width & $\unit[100]{\mu m}$ \\\hline
	Loading slot transverse acceptance angle & $\unit[\pm 4]{^\circ}$ \\\hline
	Loading slot longitudinal acceptance angle & $\unit[-45]{^\circ}$ to $\unit[+9]{^\circ}$ \\\hline
	\hline
	\xrowht{11pt}
	\textbf{Optical}  & \\\specialrule{.15em}{.2em}{.2em}
	NA skimming the surface perpendicular to trap axis & 0.11 \\\hline
	NA skimming the surface at $45^\circ$ to trap axis & 0.08 \\\hline
	NA vertically through central slot (Phoenix) & 0.25\\\hline
\end{longtable}

\pagebreak
\section{Acknowledgments}

We would like to acknowledge the Sandia National Laboratories personnel that were instrumental in developing, documenting, and characterizing these traps. 

\begin{table}[h]
	\centering
	\begin{tabular}{ccc}
	\hline	
		\textbf{Trap Packaging} 	&	 \textbf{Design and Fabrication}	&	\textbf{Design and Testing} 	 	\\ \hline
		Ray Haltli	 				&	Matthew Blain						&	Peter Maunz							\\
		Andrew Hollowell			&	Jason Dominguez 					&	Melissa Revelle						\\
		Anathea Ortega				&	Ed Heller							&	Susan Clark							\\
		Tipp Jennings				&	Corrie Sadler						&	Craig Hogle							\\
									&	Becky Loviza						&	Daniel Lobser						\\ \hhline{-~~}
		\textbf{RF Engineering}		&	John Rembetski						&	Dan Stick							\\ \hhline{-~~}	
		Christopher Nordquist		&										&	Joshua Wilson						\\ 
		Stefan Lepkowski			&  										&	Christopher Yale												
	\end{tabular}
\end{table}

\section{Attachments}
The files listed below supply additional information about the trap, trapping potentials and voltage solution generation.
\begin{longtable}{|L{0.33\textwidth}|L{0.57\textwidth}|} \hline \xrowht{14pt}
	\Large{\textbf{File Name}} 																& \Large{\textbf{Description}} \\ 
	\hline\xrowht{12pt}
	CLGA\_KDNB9B88A.pdf 				& CLGA package datasheet \\
	\hline \xrowht{12pt}
	CPGA\_KDP9F92.pdf				& CPGA package datasheet \\
	\hline \xrowht{12pt}
	H21D\_datasheet.pdf 				& Datasheet of die attachment epoxy \\
	\hline \xrowht{12pt}
	QL139401G\_W17-15.pptx	 	& Example of packaged trap documentation \\
	\hline \xrowht{12pt}
	Phoenix\_Netlist.accdb	 		& Netlist formatted in Access database \\
	\hline \xrowht{12pt}
	Phx\_Pereg\_Netlist.xlsx	 	& Netlist formatted in Excel \\
	\hline\xrowht{12pt}
	Peregrine\_Schematic.pdf	 	& Peregrine trap schematic. 10:1 scale \\
	\hline\xrowht{12pt}
	Phoenix\_Schematic.pdf	 		& Phoenix trap schematic. 10:1 scale  \\
	\hline\xrowht{12pt}
	\textbf{Phoenix-Voltage/}																& Sub-folder for Phoenix voltage control docs\\
		\hline\xrowht{12pt}
		~~~Phx\_full\_H\_axial.txt	
				& Full axial shuttling solution \\
		\hline\xrowht{12pt}
		~~~Phx\_full\_E\_x.txt		
				& Compensation solutions for \textit{x}-axis\\
		\hline\xrowht{12pt}
		~~~Phx\_full\_E\_y.txt
				& Compensation solutions for \textit{y}-axis\\
		\hline\xrowht{12pt}
		~~~Phx\_full\_E\_z.txt		
				& Compensation solutions for \textit{z}-axis\\
		\hline\xrowht{12pt}
		~~~Phx\_full\_H\_yz.txt	
				& Rotation solutions around \textit{x}-axis\\
		\hline\xrowht{12pt}
		~~~Phx\_full\_H\_xz.txt		
				& Rotation solutions around \textit{y}-axis\\
		\hline\xrowht{12pt}
		~~~Phx\_full\_H\_xy.txt		
				& Rotation solutions around \textit{z}-axis\\
		\hline\xrowht{12pt}
		~~~Phoenix.va.bin	
				& Voltage array for entire trap length\\
		\hline\xrowht{12pt}
		~~~Phoenix\_center.va.bin		
				& Voltage array for center trap region (finer steps)\\
		\hline
	\textbf{Peregrine-Voltage/}																& Sub-folder for Peregrine voltage control docs\\
		\hline\xrowht{12pt}
		~~~Prg\_full\_H\_axial.txt		
				& Full axial shuttling solution \\
		\hline\xrowht{12pt}
		~~~Prg\_full\_E\_x.txt		
				& Compensation solutions for \textit{x}-axis\\
		\hline\xrowht{12pt}
		~~~Prg\_full\_E\_y.txt		
				& Compensation solutions for \textit{y}-axis\\
		\hline\xrowht{12pt}
		~~~Prg\_full\_E\_z.txt		
				& Compensation solutions for \textit{z}-axis\\
		\hline\xrowht{12pt}
		~~~Prg\_full\_H\_yz.txt		
				& Rotation solutions around \textit{x}-axis\\
		\hline\xrowht{12pt}
		~~~Prg\_full\_H\_xz.txt		
				& Rotation solutions around \textit{y}-axis\\
		\hline\xrowht{12pt}
		~~~Prg\_full\_H\_xy.txt		
				& Rotation solutions around \textit{z}-axis\\
		\hline\xrowht{12pt}
		~~~Peregrine.va.bin		
				& Voltage array for entire trap length\\
		\hline\xrowht{12pt}
		~~~Prg\_center.va.bin		
				& Voltage array for center trap region (finer steps)\\
		\hline
	\end{longtable}

\bibliographystyle{unsrt}
\bibliography{Library}

\end{document}